\documentclass{IEEEtran}

\usepackage{enumitem,kantlipsum}
\usepackage{graphicx}
\usepackage{epstopdf}
\usepackage{amssymb}
\usepackage{amsmath}
\usepackage{subfigure}
\usepackage{epsfig}
\usepackage{color}
\usepackage{enumitem}
\usepackage{float}
\usepackage{soul}
\usepackage{wrapfig}
\usepackage{arydshln}
\usepackage{tablefootnote}

\usepackage{amsthm}

\def\N{\mathcal{N}}

\def\0{{\bf 0}}
\def\1{{\bf 1}}

\def\beq{\begin{equation*}}
    \def\eeq{\end{equation*}}
\def\bql{\begin{equation}}
    \def\eql{\end{equation}}
\def\bqn{\begin{eqnarray*}}
    \def\eqn{\end{eqnarray*}}
\def\bnl{\begin{eqnarray}}
    \def\enl{\end{eqnarray}}
\def\bma{\begin{bmatrix}}
    \def\ema{\end{bmatrix}}
\def\bmx{\begin{matrix}}
    \def\emx{\end{matrix}}
\def\ben{\begin{enumerate}}
    \def\een{\end{enumerate}}
\def\bit{\begin{itemize}}
    \def\eit{\end{itemize}}
\def\bei{\begin{itemize}}
    \def\eei{\end{itemize}}
\def\bet{\begin{tabular}}
    \def\eet{\end{tabular}}

\newcommand{\ba}{\mathbf{a}}

\newcommand{\be}{\mathbf{e}}

\usepackage[dvipsnames]{xcolor}

\def\1{{\bf1}}

\def\b{{\beta}}
\def\a{\alpha}

\def\bit{\begin{itemize}}
\def\eit{\end{itemize}}

\def\be{\begin{equation}}
\def\ee{\end{equation}}
\def\ba{\begin{eqnarray}}
\def\ea{\end{eqnarray}}

\def\bes{\begin{equation*}}
\def\ees{\end{equation*}}
\def\bas{\begin{eqnarray*}}
\def\eas{\end{eqnarray*}}

\usepackage{url}
\usepackage{verbatim}

\newtheorem{Remark 1}{Remark}
\newtheorem{Remark 2}[Remark 1]{Remark}
\newtheorem{Remark 3}[Remark 1]{Remark}
\newtheorem{Remark 4}[Remark 1]{Remark}
\newtheorem{Remark 5}[Remark 1]{Remark}
\newtheorem{Remark 6}[Remark 1]{Remark}
\newtheorem{Remark 7}[Remark 1]{Remark}

\newtheorem{Lemma 1}{Lemma}
\newtheorem{Lemma 2}[Lemma 1]{Lemma}
\newtheorem{Lemma 3}[Lemma 1]{Lemma}
\newtheorem{Lemma 4}[Lemma 1]{Lemma}
\newtheorem{Lemma 5}[Lemma 1]{Lemma}
\newtheorem{Lemma 6}[Lemma 1]{Lemma}
\newtheorem{Lemma 7}[Lemma 1]{Lemma}

\newtheorem{Assumption 1}{Assumption}
\newtheorem{Assumption 2}[Assumption 1]{Assumption}
\newtheorem{Assumption 3}[Assumption 1]{Assumption}
\newtheorem{Assumption 4}[Assumption 1]{Assumption}
\newtheorem{Definition 1}{Definition}
\newtheorem{Theorem 1}{Theorem}
\newtheorem{Theorem 2}[Theorem 1]{Theorem}
\newtheorem{Theorem 3}[Theorem 1]{Theorem}
\newtheorem{Theorem 4}[Theorem 1]{Theorem}
\newtheorem{Theorem 5}[Theorem 1]{Theorem}
\newtheorem{Theorem 6}[Theorem 1]{Theorem}
\newtheorem{Theorem 7}[Theorem 1]{Theorem}
\newtheorem{Theorem 8}[Theorem 1]{Theorem}
\newtheorem{Theorem 9}[Theorem 1]{Theorem}
\newtheorem{Theorem 10}[Theorem 1]{Theorem}

\newtheorem{Proposition 1}{Proposition}

\title{\LARGE \bf
 Ensure  Differential Privacy and  Convergence Accuracy in   Consensus Tracking and Aggregative Games with Coupling Constraints}

\author{Yongqiang Wang
\thanks{   The work   was supported in part by the National Science Foundation under Grants   ECCS-1912702, CCF-2106293, CCF-2215088, and
CNS-2219487.}
\thanks{Yongqiang Wang is with the Department of Electrical and Computer Engineering, Clemson University, Clemson, SC 29634, USA
{\tt\small{yongqiw}@clemson.edu}
}%

}

\begin{document}

\maketitle
\thispagestyle{empty}
\pagestyle{empty}

\begin{abstract}
We address differential privacy for fully distributed aggregative games with shared coupling constraints. By co-designing the generalized Nash equilibrium (GNE) seeking mechanism and the differential-privacy noise injection mechanism, we propose the first GNE seeking algorithm that can ensure both provable convergence to the GNE and rigorous $\epsilon$-differential privacy, even with the number of iterations tending to infinity. As a basis of the co-design, we  propose a new consensus-tracking algorithm that can achieve rigorous $\epsilon$-differential privacy while maintaining accurate tracking performance, which, to our knowledge,  has not been achieved before.   To facilitate the convergence analysis, we also establish a general convergence result  for stochastically-perturbed nonstationary fixed-point iteration processes, which lie at the core of numerous optimization and variational problems. Numerical simulation results   confirm the effectiveness of the proposed approach.
\end{abstract}

\section{Introduction}

In recent years,   Nash equilibrium seeking in aggregative games under shared coupling constraints is gaining increased traction \cite{pavel2019distributed,yi2019operator,belgioioso2020distributed}. Compared with the standard Nash equilibrium seeking problem where the feasible decision sets of individual players (also called agents) are independent of each other, the incorporation of  shared coupling constraints in Nash equilibrium seeking, usually called the generalized Nash equilibrium (GNE) model,  captures important characteristics of noncooperative games under limited network resources. To date, the GNE seeking problem   has found vast applications in economics and various engineering domains, with typical examples including power grids \cite{saad2012game,zhu2016distributed,ye2016game}, optical networks \cite{pan2009games}, communication networks \cite{alpcan2005distributed,yin2011nash,yong2008fault}, plug-in electric cars \cite{ma2011decentralized,ma2016efficient}, and mobile sensor networks \cite{stankovic2011distributed,wang2013statistical,wang2011influences}.
In many of these applications, due to practical constraints, no central coordinator/mediator exists  to collect and disperse information, and individual players can only access or observe the decisions of their immediate neighbors. Compared with the traditional case where every player can access all its competitors' decisions to precisely evaluate its cost function, which is called the {\it full-decision information} setting, the case where each player can only access its neighbors' decisions is  called the {\it partial-decision information} setting, and is more challenging in that individual players  lack  information to compute their cost functions or gradients.

To account for the lack of information in the partial-decision information setting, players have to exchange information among local neighbors
to estimate  global information necessary for GNE seeking.  Since the seminal work in  \cite{koshal2016distributed,salehisadaghiani2016distributed}, significant inroads have been make in games with  partial-decision information, both  for Nash equilibrium seeking without coupling constraints (see, e.g., \cite{tatarenko2020geometric,gadjov2018passivity,ye2017distributed,salehisadaghiani2019distributed}) and GNE seeking (see, e.g., \cite{pavel2019distributed,belgioioso2020distributed,bianchi2022fast}). A key technique underlying these fully distributed algorithms is consensus tracking \cite{zhu2010discrete} (also called dynamic average consensus), which allows a group of agents to locally track the average of their reference inputs  while each agent  only has access to its own individual reference input. However, the employment of consensus tracking in GNE seeking also entails   explicit sharing of (estimated) decisions in every iteration, which is problematic when involved information is sensitive \cite{zhang2022privacy_dynamic,wang2022quantization,wang2022decentralized_inherent}.  Given the   noncooperative relationship between players, individual players' privacy should be protected to avoid exploitation by  others.  For example, in Nash-Cournot games, individual players' cost functions could be market sensitive, and should not be disclosed to competitors \cite{rassenti2000adaptation}. In fact, in many scenarios, privacy preservation is mandated by legislation. For example, in road routing games \cite{dong2015differential},  California Privacy Rights Act forbids disclosing drivers' spatiotemporal information, which, otherwise, can reveal a person's activities \cite{gerhart2020proposition}. Moreover, privacy preservation is also a crucial step to  encourage  participation in cooperative policies \cite{shakarami2022distributed}.

To address the privacy issue in   games, several approaches have been proposed for the full-decision information setting where a coordinator exists (see, e.g., \cite{lu2015game,cummings2015privacy,shilov2021privacy}).   Recently, results have also emerged to protect privacy in the partial-decision information setting (see, e.g., \cite{shakarami2022distributed,gade2020privatizing}). However, these approaches are restricted  in that they either require the communication graph to satisfy certain properties, or can only protect the   cost function value  from being uniquely identifiable.   As differential privacy (DP) has become the de facto standard for privacy protection due to its  strong resilience against arbitrary post-processing and  auxiliary information \cite{dwork2006calibrating},  recently  \cite{ye2021differentially} and \cite{wang2022differentially} achieve DP in   aggregative games at the cost of losing   convergence to the exact Nash equilibrium. By co-designing the Nash-equilibrium seeking mechanism and the DP-noise injection mechanism, our prior results \cite{wang2022aggregate,wang2022ensuring} successfully achieve both $\epsilon$-DP and accurate convergence in coupling-constraint free Nash-equilibrium seeking.      However, these results   are inapplicable in GNE seeking, where the shared coupling constraints increase attack surfaces, and hence, pose additional challenges to privacy protection. To   our knowledge,  privacy protection  for GNE seeking is still an open problem in the partial-decision information setting.

In this paper, we propose a fully distributed GNE seeking algorithm that  can ensure both accurate convergence
and rigorous $\epsilon$-DP, with the cumulative privacy budget guaranteed to be finite even when the number of iterations tends to infinity.
Our  approach is motivated by the observation that DP-noises enter the algorithm
through inter-player interaction, and hence, their influence on convergence accuracy can be attenuated
by gradually weakening inter-player interaction, which would  become unnecessary anyway
after convergence.   To ensure that every player can still estimate global information necessary for GNE seeking,
  we judiciously design the weakening factor sequence for inter-player interaction and the stepsize sequence,  which enables the achievement of
    accurate convergence  even in the presence of DP-noises. We
  prove that the algorithm is $\epsilon$-differentially private
with a finite cumulative privacy budget, even on the infinite time horizon. Given that a key component of our GNE seeking algorithm is  consensus tracking   but the conventional consensus-tracking algorithm will lead to cumulative and exploding variance in the presence of persistent noise \cite{wang2022gradient}, we first propose a new robust consensus-tracking algorithm that can ensure both provable convergence and $\epsilon$-DP under persistent DP-noise.

The main contributions are summarized as follows:
\begin{enumerate}[wide, labelwidth=!,labelindent=8pt]
  \item  The proposed new consensus-tracking approach  can ensure  both provable convergence accuracy and $\epsilon$-DP, a goal that has not been achieved before to the best of our knowledge.
  \item The proposed   fully distributed GNE seeking algorithm, to  our knowledge,  is the first  to achieve rigorous $\epsilon$-DP  in GNE seeking under  partial-decision information. {Note that compared with our prior work on differentially private NE seeking  without coupling constraints \cite{wang2022aggregate,wang2022ensuring}, not only is the convergence analysis  completely different due to the need to use dual variables to address shared coupling constraints (we use the operator-theoretic approach to facilitate the analysis here, which is not used in \cite{wang2022aggregate,wang2022ensuring}), the DP design is also much more complicated due to the need to share more information (both the primal and dual variables) in GNE seeking. In fact, the shared coupling constraints in GNE problems create additional attack surfaces (the additional shared variables provide   attackers with more data to infer sensitive information) and thus challenges in privacy protection.}
  \item  Besides achieving rigorous $\epsilon$-DP, the approach can simultaneously ensure  provable convergence to the GNE,  which is in sharp contrast to existing DP solutions for coupling-constraint free aggregative games   (e.g., \cite{ye2021differentially,wang2022differentially}) that have to trade provable convergence for DP.
  \item  To facilitate convergence analysis under DP-noises, we establish  convergence results for stochastically-perturbed nonstationary fixed-point iteration processes. Given that fixed-point iteration  processes lie at the core of many optimization and variational problems, the results are expected to have broad ramifications.
  \item  Even without considering privacy protection, our proof techniques are fundamentally different from existing counterparts and are  of independent interest. More specifically, existing proof techniques (in, e.g.,  \cite{belgioioso2020distributed,koshal2016distributed,wang2022differentially,parise2019distributed,gadjov2020single,zhu2022asynchronous,liang2017distributed,lei2022distributed}) for partial-decision information games rely on the geometric decreasing of consensus errors among the players.
      In the proposed approach, the diminishing  interaction  makes it impossible to have such  geometric decreasing of consensus errors, which entails new proof techniques.
  \end{enumerate}

The organization of the paper is as follows.
Sec.~\ref{sec-problem} gives the problem formulation
and some results for a later use. Sec. \ref{se:algorithm1} presents a new consensus-tracking algorithm. This section also proves that  the algorithm can ensure both provable accurate convergence and rigorous $\epsilon$-DP with a finite cumulative privacy budget. Sec. \ref{se:algorithm_2} proposes a fully distributed GNE seeking algorithm, while its convergence and privacy analysis are presented in Sec. \ref{se:convergence_algo2} and Sec. \ref{se:DP_algo2}, respectively.  Sec. \ref{se:simulation} presents numerical simulation results to confirm the obtained results. Finally, Sec. \ref{se:conclusions} concludes the paper.

{\bf Notations:}
 We use $\mathbb{R}^d$ to denote the Euclidean space of
dimension $d$ and $\mathbb{R}^d_{+}$  the set of all nonnegative vectors in  $\mathbb{R}^d$. We write $I_d$ for the identity matrix of dimension $d$,
and ${\bf 1}_d$ for  the $d$-dimensional  column vector with all
entries equal to 1; in both cases  we suppress the dimension when it is
clear from the context. 
A vector is viewed as a column
vector, and for  a
vector $x$, $[x]_i$ denotes its $i$th element.
  We use $\langle\cdot,\cdot\rangle$ to denote the inner product and
 $\|x\|$ for the standard Euclidean norm of a vector $x$. We use $\|x\|_1$ to represent the $L_1$ norm of a vector $x$.
We write $\|A\|$ for the matrix norm induced by the vector norm $\|\cdot\|$.
 $A^T$ denotes the transpose of a matrix $A$.
 Given vectors $x_1,\cdots,x_m$, we define $x={\rm col}(x_1,\cdots,x_m)=[x_1^T,\cdots,x_m^T]^T$, $x_{-i}={\rm col}(x_1,\cdots,x_{i-1},x_{i+1},\cdots,x_m)$, and $\bar{x}=\frac{\sum_{i=1}^{m}x_i}{m}$. We define $[m]\triangleq \{1,\,2,\,\cdots,m\}$.
  Often, we abbreviate {\it almost surely} by {\it a.s}.
\def\as{{\it a.s.\ }} Given  sets $\Omega_1,\cdots,\Omega_m$, we use $\Omega_1\times \cdots\times \Omega_m$ or $\Pi_{i=1}^m\Omega_i$ to represent the Cartesian product of these sets. {For a variable $r^k$, we use $\{r^k\}$ to denote the sequence of values of $r^k$ for all $k\geq 0$}.


\section{Problem Formulation and Preliminaries}\label{sec-problem}
\subsection{Monotone Operators}
All of the following are adopted from   \cite{bauschke2011convex}.
Let $\mathcal{X}$ and $\mathcal{Y}$ be non-empty sets, and $2^{\mathcal{Y}}$ be the family of all subsets of $\mathcal{Y}$. An operator $T:\mathcal{X}\rightarrow \mathcal{Y}$ is a mapping that maps every point $x\in \mathcal{X}$ to a point $T(x)$ in $\mathcal{Y}$. Thus, the notion $T:\mathcal{X} \rightarrow 2^{\mathcal{Y}}$ means that $T$ is a set-valued operator from $\mathcal{X}$ to $\mathcal{Y}$. $\rm Id$ denotes the identity operator. The domain of an operator $T$ is represented by ${\rm dom}T=\{x|T(x)\neq \emptyset\}$ where $\emptyset$ denotes the empty set. The range of an operator $T$ is defined by ${\rm ran}T=\{u|\exists x, u\in T(x)\}$. The graph of $T$ is defined as ${\rm gra}T=\{(x,u)|u\in T(x)\}$, and the inverse of an operator is defined using graph as ${\rm gra}T^{-1}={(u,x)|u\in T(x)}$. The zero set of $T$ is defined as ${\rm zer}T=\{x|0\in T(x)\}$.

We  consider monotone operators, i.e., operators satisfying $\langle x-y,u-v\rangle\geq 0$ for all $(x,u)$ and $(y,v)$ in ${\rm gra}T$. An operator $T$ is maximally monotone if ${\rm gra}T$ is not contained in the graph of any other monotone operator.  $T$ is $\eta$-strongly monotone, with $\eta>0$, if $\langle x-y,u-v\rangle\geq \eta\|x-y\|^2$ for all $(x,u)$ and $(y,v)$ in ${\rm gra}T$. A  set-valued operator $T:\mathbb{R}^d\rightarrow 2^{\mathbb{R}^d}$ is called restricted-strictly monotone with respect to $\mathcal{U}\subset\mathbb{R}^d$ if $\langle x^\ast-x,u^\ast-u\rangle>0$ for all $u^\ast\in\mathcal{U}$, $u\in\mathbb{R}^d\setminus \mathcal{U}$, $x^\ast\in T(u^\ast)$, and $x\in T(u)$. For an operator $T:\mathbb{R}^d\rightarrow 2^{\mathbb{R}^d}$, we use $({\rm Id}+T)^{-1}$ to represent its resolvent, which is single valued and has domain equal to $\mathbb{R}^d$ if $T$ is maximally monotone. For  two operators $T_1$ and $T_2$, their composition is denoted as $T_1\circ T_2$, and their sum  $T_1+T_2$ is defined as ${\rm gra}(T_1+T_2)=\{(x,y+z)|(x,y)\in {\rm gra}A, (x,z)\in {\rm gra}T_2\}$. If $T_2$ is single valued, then we always have ${\rm zer}(T_1+T_2)={\rm Fix}( ({\rm Id}+T_1 )^{-1}\circ ({\rm Id}-T_2) )$, where ${\rm Fix}(T)$ denotes the fixed point set  $\{x \in {\rm dom}T|T(x)=x\}$.

The subdifferential $\partial f$ of a proper lower semicontinuous convex function $f:x \rightarrow 2^{\mathbb{R}^d}$ is defined as $x\rightarrow \{g\in \mathbb{R}^d|f(y)\geq f(x)+\langle g,y-x\rangle,\forall y\in {\rm dom}f\}$.  We define the indicator function of $\Omega$ as $\iota_{\Omega}(x)$ ($\iota_{\Omega}(x)=0$ for $x\in \Omega$ and $\iota_{\Omega}(x)=\infty$ for $x\notin \Omega$). 
$\partial \iota_{\Omega}$ is usually called the normal cone operator of $\Omega$, and is denoted as $N_{\Omega}$.

Operator $T$ is nonexpansive if it satisfies $\|T(x)-T(y)\|\leq \|x-y\|$ for all $x$ and $y$ in ${\rm dom}T$. It is   $\alpha$-averaged for $0<\alpha<1$  if there exists a nonexpansive operator $R$ such that $T=(1-\alpha){\rm Id}+\alpha R$ holds.   $T$ being $\alpha$-averaged is equivalent to $\|T(x)-T(y)\|^2\leq \|x-y\|^2-\frac{1-\alpha}{\alpha}\|(x-y)-(Tx-Ty)\|^2$ for all $x,y\in {\rm dom}T$ \cite{bauschke2011convex}.
$T$ is called ${\beta}$-cocoercive for ${\beta}>0$ if ${\beta} T$ is $\frac{1}{2}$-averaged, i.e., $\|{\beta} T(x)-{\beta} T(y)\|^2 \leq  \langle x-y, Tx-Ty\rangle$. If $f$ is convex differentiable, with its gradient $\nabla f$ being  $\theta$-Lipschitz, then $\nabla f$ is $\frac{1}{\theta}$-cocoercive.

\subsection{On GNE Seeking}
We consider a GNE problem  among a set of $m$ players (agents) $[m]=\{1,\,\cdots,m\}$. We index the players by $1,\,2,\,\cdots,m$. Player $i$ is characterized by a strategy set $\Omega_i\subseteq \mathbb{R}^d$ and a cost function $J_i(x_i,\bar{x})$, which is dependent on both its own decision $x_i{\in \Omega_i \subseteq\mathbb{R}^{d}}$ and the aggregate of all players' decisions
$\bar{x}\triangleq \frac{\sum_{i=1}^{m} {x_i}}{m}$.  
Moreover, the decisions of all players must satisfy a shared  affine coupling constraint
\begin{equation}{\label{eq:constraint}}
\textstyle\sum_{i=1}^{m} C_ix_i\leq \sum_{i=1}^{m}c_i,
\end{equation}
where $C_i\in\mathbb{R}^{n\times d}$ and $c_i\in\mathbb{R}^{n}$ are local parameters only known to player $i$. It is worth noting that such affine coupling constraints arise in various game scenarios involving upper or lower limits of shared resources \cite{saad2012game,ma2011decentralized,li2015demand,barrera2014dynamic}, and have been a common assumption in the literature of noncooperative games \cite{pavel2019distributed,yi2019operator,belgioioso2020distributed,parise2019distributed,paccagnan2018nash}. For notational simplicity, we define $C\triangleq [C_1,\cdots,C_m]\in\mathbb{R}^{n\times md}$ and $c\triangleq \sum_{i=1}^{m}c_i\in\mathbb{R}^n$.

With these notations, we can formalize   the game that player $i$ faces as the following parameterized optimization problem:
\begin{equation}\label{eq:formulation}
\begin{aligned}
& \qquad\qquad \qquad\min_{ x_i} J_i(x_i,\bar{x})\\
&  {\rm s.t.}\quad x_i\in \Omega_i\:\: {\rm and}\:\: C_ix_i-c_i\leq \textstyle\sum_{j\neq i,j\in[m]}c_j-C_jx_j.
\end{aligned}
\end{equation}
The constraint set $\Omega_i$, the function $J_i(\cdot,\cdot)$, and the coupling matrices $C_i$ as well as $c_i$ are  known to player $i$ only.

\begin{Assumption 1}\label{ass:Kset}
  For every player $i\in[m]$, $J_i(x_i,\bar{x})$ is a differentiable convex function with respect to $x_i$ given any fixed $x_{-i}$. $\Omega_i\subset\mathbb{R}^d$ is nonempty, compact, and convex.  Moreover, the global feasible set
  \begin{equation}\label{eq:global_feasible_set}
  K\triangleq \Pi_{i=1}^m\Omega_i\bigcap\left\{x\in\mathbb{R}^{md}|\textstyle\sum_{i=1}^{m} C_ix_i\leq \textstyle\sum_{i=1}^{m}c_i\right\}
  \end{equation}
   is nonempty and satisfies  Slater’s constraint qualification.
\end{Assumption 1}

A GNE of game (\ref{eq:formulation})  is a collective strategy $x^{\ast}={\rm col}(x_1^\ast,\cdots,x_m^\ast)$ such that the following  relationship holds for any $i\in[m]$ and $x\in K$:
\[
\textstyle J_i(x_i^\ast,\bar{x}^\ast)\leq J_i(z,\frac{z}{m} +\frac{\sum_{j\neq i,j\in[m]}x_j^\ast}{m}),\: {\rm s.t.} (z,x_{-i}^\ast)\in K.
\]

Inspired by the recent success of the operator-theoretic approach to GNE seeking \cite{pavel2019distributed,yi2019operator,belgioioso2020distributed,belgioioso2022distributed}, we will employ a monotone-operator based approach \cite{bauschke2011convex} to treat the convergence to GNE under DP-noises. More specifically, according to \cite{pavel2019distributed,yi2019operator,belgioioso2020distributed}, a set of strategies $x^\ast$ is a GNE of the game in (\ref{eq:formulation}) if and only if the following conditions are satisfied for { some $\lambda_1^{\ast},\cdots, \lambda_m^{\ast}\in\mathbb{R}^{n}_{+}$}:
\begin{equation}\label{eq:KKT}
\hspace{-0.3cm}\forall i\in[m],\quad   \left\{\begin{array}{l}0 \in \bigtriangledown_{x_i}J(x_i^\ast,\bar{x}^\ast)+N_{\Omega_i}(x_i^\ast)+C_i^T\lambda_i^\ast\\
   0\leq \lambda_i^\ast \:\:\bot-(C x^\ast-c)\geq 0 \end{array}\right.,
\end{equation}
{where the symbol $\bot$ means  perpendicular (namely, given $u,v\in\mathbb{R}^n$, $u\bot v$ means $u^Tv=0$).}

Similar to \cite{pavel2019distributed,yi2019operator,belgioioso2020distributed}, we are interested in a type of GNE called variational GNE, which has the economic interpretation of no price discrimination \cite{kulkarni2012variational}, and corresponds to a solution set of (\ref{eq:KKT}) with equal dual variables $\lambda_1^\ast=\cdots=\lambda_m^\ast$ \cite{pavel2019distributed,facchinei2003finite}. Recently, \cite{belgioioso2020distributed}
shows that the variational GNE corresponds to the zero set of the following set-valued mapping:
\begin{equation}\label{eq:T}
 T:\left[\begin{array}{c}x\\\lambda\end{array}\right]\mapsto \left[\begin{array}{c}N_{\Omega}(x)+F(x)+\frac{C_f^T\lambda}{m}\\N_{\mathbb{R}^{mn}_{+}}(\lambda)+\Pi_f\lambda-\frac{C_fx-c_f}{m}\end{array}\right],
\end{equation}
where $x={\rm col}(x_1,\cdots,x_m)$, $\lambda={\rm col}(\lambda_1,\cdots,\lambda_m)$, $C_f={\bf 1}_m\otimes C$, $c_f={\bf 1}_m\otimes c$,
$\Pi_f\triangleq (I_m-\frac{{\bf 1}{\bf 1}^T}{m})\otimes I_n$, and $F(x)$ is the pseudogradient of the game defined as
\begin{equation}\label{eq:F(x)}
  F(x)={\rm col}(\nabla_{x_1}J(x_1,\bar{x}),\cdots,\nabla_{x_m}J(x_m,\bar{x})).
\end{equation}

\begin{Assumption 1}\label{ass:cocoersive}
  $F(\cdot )$ in (\ref{eq:F(x)}) is $\mu$-{ strongly monotone} over $\mathbb{R}^{md}$.
\end{Assumption 1}

Using the definition of $F(\cdot)$, we can further define
$
F_i(v,u)\triangleq {(\frac{\partial}{\partial z_1}J_{i}(z_1,z_2) +\frac{1}{m}\frac{\partial}{\partial z_2}J_{i}(z_1,z_2) )|_{z_1=v,z_2=u} }
$
and
$
F(v,u)\triangleq {\rm col}(F_1(v,u),\,\cdots,\,F_m(v,u))$.
It can be verified that $F(x,{\bf 1}\otimes  \bar{x})$ is equal to $F(x)$ defined in (\ref{eq:F(x)}).

Moreover, following \cite{pavel2019distributed,belgioioso2020distributed,koshal2016distributed},
we  make the following assumption on the mapping $F(\cdot,\cdot)$:
\begin{Assumption 1}\label{ass:Lipschitz}
  The exists a constant $\tilde{L}$ such that $F(\cdot,\cdot)$ satisfies $\|F(u_1,v_1)-F(u_2,v_2)\|\leq \tilde{L}\left\|\left[\begin{array}{c}u_1\\v_1\end{array}\right]-\left[\begin{array}{c}u_2\\v_2\end{array}\right]\right\|$ for all $u_1,u_2 { \in\mathbb{R}^{md}}$ and $v_1,v_2\in\mathbb{R}^{md}$.
\end{Assumption 1}

{ By combining Assumption \ref{ass:cocoersive} and Assumption (\ref{ass:Lipschitz}), we have that $ F $ is $\frac{\mu}{\tilde{L}^2}$-cocoercive over $\mathbb{R}^{md}$ \cite{bauschke2011convex}.}

Using operator splitting \cite{bauschke2011convex}, the mapping $T$ in (\ref{eq:T}) can be split into  the sum of two mappings:
 \begin{equation}\label{eq:T_1+T_2}
   T_1\hspace{-0.1cm}: \left[\hspace{-0.1cm}\begin{array}{c}x\\\lambda\end{array}\hspace{-0.1cm}\right]\hspace{-0.1cm} \mapsto \hspace{-0.1cm} \left[\hspace{-0.1cm}\begin{array}{c}F(x)\\\Pi_f\lambda+\frac{c_f}{m}\end{array}\hspace{-0.1cm}\right],\:
   T_2\hspace{-0.1cm}:\left[\hspace{-0.1cm}\begin{array}{c}x\\\lambda\end{array}\hspace{-0.1cm}\right]\hspace{-0.1cm} \mapsto \hspace{-0.1cm} \left[\hspace{-0.1cm}\begin{array}{c}N_\Omega(x)+\frac{C_f^T\lambda}{N}\\N_{\mathbb{R}^{mn}_{+}}(\lambda)-\frac{C_f x}{m} \end{array}\hspace{-0.1cm}\right].
 \end{equation}
 It has been proven in \cite{belgioioso2020distributed}
that the operator $T_2$ is maximally monotone, and that the operator $T_1$ is $\delta$-cocoercive ($0<\delta\leq \min\{1,{\frac{\mu}{\tilde{L}^2}}\}$ with $\mu$ from   Assumption \ref{ass:cocoersive} {and $\tilde{L}$ from Assumption \ref{ass:Lipschitz}}) and {restricted-}strictly monotone { with respect to $\mathbb{R}^{md}\times E^{||}$ where $E^{||}$ denotes the consensus subspace of dual variables}. Furthermore, \cite{belgioioso2020distributed} also showed that when $T$ is restricted-strictly monotone { with respect to $\mathbb{R}^{md}\times E^{||}$}, the variational GNE of the game in (\ref{eq:formulation}) is fully characterized by the zeros of the operator $T$:
\begin{Lemma 1}\cite{belgioioso2020distributed}
  Under Assumption \ref{ass:Kset}, 1) if ${\rm zer}(T)\neq \emptyset$ and ${
  \rm col}(x^{\ast},\lambda^{\ast})\in{\rm zer}(T)$, then $x^{\ast}$ is a variational GNE and $\lambda_1^{\ast}=\lambda_2^{\ast}=\cdots=\lambda_m^{\ast}\in\mathbb{R}^{n}_{+}$; { 2)  if a variational GNE
exists, then $zer(T)$ is nonempty.}
\end{Lemma 1}

We consider distributed algorithms for  the game in (\ref{eq:formulation}), where no player has  direct access to the average decision $\bar{x}$. Instead, each player has to  locally estimate $\bar{x}$  through   interactions with its neighbors. We describe the local interaction using a weight matrix
$L=\{L_{ij}\}$, where $L_{ij}>0$ if  { player  $j$ can   directly communicate with player $i$},
and $L_{ij}=0$ otherwise. For   player $i\in[m]$,
its  neighbor set
$\mathbb{N}_i$ is  the collection of players $j$ such that $L_{ij}>0$.
We define $L_{ii}\triangleq-\sum_{j\in\mathbb{N}_i}L_{ij}$  for all $i\in [m]$.
We make the following assumption on $L$:

\begin{Assumption 1}\label{as:L}
 The matrix  $L=\{w_{ij}\}\in \mathbb{R}^{m\times m}$ is symmetric and satisfies
    ${\bf 1}^TL={\bf
  0}^T$, $L{\bf 1}={\bf
  0}$, and $ \|I+L-\frac{{\bf 1}{\bf 1}^T}{m}\|<1$.
\end{Assumption 1}


We also need  the following lemma for convergence analysis:
\begin{Lemma 2}\cite{wang2023tailoring}\label{Lemma-polyak}
Let $\{v^k\}$, $\{\a^k\}$, $\{p^k\}$ be random nonnegative scalar sequences, and
$\{q^k\}$ be a deterministic nonnegative scalar sequence satisfying
$\sum_{k=0}^\infty \a^k<\infty$  {\it a.s.},
$\sum_{k=0}^\infty q^k=\infty$, $\sum_{k=0}^\infty p^k<\infty$ {\it a.s.} If there exists a $k_0\geq 0$ such that
\[
\mathbb{E}\left[v^{k+1}|\mathcal{F}^k\right]\le(1+\a^k-q^k) v^k +p^k,\quad \forall k\geq k_0\quad\as
\]
where $\mathcal{F}^k=\{v^\ell,\a^\ell,p^\ell; 0\le \ell\le k\}$,
then, $\sum_{k=0}^{\infty}q^k v^k<\infty$ and
$\lim_{k\to\infty} v^k=0$ hold almost surely.
\end{Lemma 2}
\begin{Lemma 4}\cite{wang2023tailoring}\label{le:chung}
Let $\{v^k\}$ be a nonnegative sequence, and  $\{\a^k\}$  and $\{\b^k\}$ be positive non-increasing sequences satisfying $\sum_{k=0}^{\infty}\a^k=\infty$, $\lim_{k\rightarrow \infty} \a^k =0$, and $\lim_{k\rightarrow \infty}\frac{\b^k}{\a^k}=0$. If there exists a $k_0\geq 0$ such that $ v^{k+1} \le(1-\a^k) v^k +\b^k$ holds for all $k\geq k_0$,
 then  $v^k\leq C \frac{\b^k}{\a^k}$ holds for all $k$, where $C$ is some constant.
\end{Lemma 4}

\subsection{On Differential Privacy}
We use the notion of $\epsilon$-DP for continuous bit streams \cite{dwork2010differential}, which has recently been applied to distributed optimization  (see \cite{Huang15} as well as our own work \cite{wang2023tailoring}). To enable $\epsilon$-DP, we add Laplace noise to all shared messages. For a constant $\nu>0$, we use ${\rm Lap}(\nu)$ to denote a Laplace distribution of a scalar random variable with the probability density function $x\mapsto\frac{1}{2\nu}e^{-\frac{|x|}{\nu}}$.  ${\rm Lap}(\nu)$'s mean is   zero   and its  variance is $2\nu^2$.
To facilitate DP analysis, we represent the   game $\mathcal{P}$ in (\ref{eq:formulation}) by three parameters ($K, \mathcal{J}, L$), where
 $K$ defined in (\ref{eq:global_feasible_set}) is the domain of decision variables,     $ \mathcal{J} \triangleq\{ J_1,\,\cdots,J_m\}$, and $ L$ is the   weight matrix. Then we define adjacent  games as follows:

\begin{Definition 1}\label{de:adjacency}
Two games $\mathcal{P}=( K, \mathcal{J}, L)$ and $\mathcal{P}'=(K', \mathcal{J}', L')$ are adjacent if the following conditions hold:
\begin{itemize}
\item $K=K'$  and $L=L'$, i.e., the domains of decision variables   and the  weight matrices are identical;
\item there exists an $i\in[m]$ such that $J_i\neq J_i'$ but $J_j=J_j'$ for all $j\in[m],\,j\neq i$;
\item  the different functions $J_i$ and $J'_i$  have similar behaviors   around $x^\ast$, the GNE of $\mathcal{P}$. More specifically,  there exits some $\delta>0$  such that for all $v={\rm  col}(v_1,\cdots,v_m)$ and $v'={\rm col}(v'_1,\cdots,v'_m)$ in  $B_\delta(x^\ast)\triangleq\{u:u\in\mathbb{R}^{md}, \|u-x^\ast\|<\delta\}$, we have $\nabla F_i(v_i,\bar{v})=\nabla F'_i(v_i',\bar{v}')$ where $\bar{v}=\frac{\sum_{i=1}^{m}v_i}{m}$ and $\bar{v}'=\frac{\sum_{i=1}^{m}v'_i}{m}$.
\end{itemize}
\end{Definition 1}

 Definition \ref{de:adjacency} implies that two GNE problems are adjacent if  one player changes its cost function   while all other game characteristics are identical.

  \begin{Remark 1}
 In Definition \ref{de:adjacency}, since  the change  from $J_i$ to $J'_i$ in the second condition  can be arbitrary,  the third condition is added to restrict the change magnitude. It is necessary to enabling rigorous $\epsilon$-DP while maintaining accurate convergence.  This is because  DP aims to make observations statistically indistinguishable while accurate convergence means that the state will stop changing and remain time-invariant after a transient period. Hence,  to make the observations of $\mathcal{P}$ and $\mathcal{P}'$ the same after their states converge and remain  at  their respective converging points,  we have to require $\mathcal{P}$ and $\mathcal{P}'$ to have identical converging points.
\end{Remark 1}

 Given a distributed GNE seeking algorithm, we represent an execution of such an algorithm as $\mathcal{A}$, which is an infinite sequence of the iteration variable $\vartheta$, i.e., $\mathcal{A}=\{\vartheta^0,\vartheta^1,\cdots\}$. We consider adversaries that can observe all communicated messages among the players. Therefore, the observation part of an execution is the infinite sequence of shared messages, which is denoted as $\mathcal{O}$. Given a distributed GNE  problem $\mathcal{P}$ and an initial state $\vartheta^0$, we define the observation mapping as $\mathcal{R}_{\mathcal{P},\vartheta^0}(\mathcal{A})\triangleq \mathcal{O}$. Given a GNE  problem $\mathcal{P}$, observation sequence $\mathcal{O}$, and an initial state $\vartheta^0$,  $\mathcal{R}_{\mathcal{P},\vartheta^0}^{-1}(\mathcal{O})$ is the set of executions $\mathcal{A}$ that can generate the observation $\mathcal{O}$.
 \begin{Definition 1}
   ($\epsilon$-differential privacy, adapted from \cite{Huang15}). For a given $\epsilon>0$, an iterative distributed algorithm  is $\epsilon$-differentially private if for any two adjacent { network  games} $\mathcal{P}$ and $\mathcal{P}'$, any set of observation sequences $\mathcal{O}_s\subseteq\mathbb{O}$ (with $\mathbb{O}$ denoting the set of all possible observation sequences), and any initial state ${\vartheta}^0$, the following relationship always holds
    \begin{equation}\label{eq:DP_definition1}
        \mathbb{P}[\mathcal{R}_{\mathcal{P},\vartheta^0}\in\mathcal{O}_s]\leq e^\epsilon\mathbb{P}[\mathcal{R}_{\mathcal{P}',\vartheta^0}\in\mathcal{O}_s],
    \end{equation}
    with the probability $\mathbb{P}$  taken over  randomness in  iterations.
 \end{Definition 1}

 {Since the observation sequence  in Definition 2 involves observation values in multiple iterations, the privacy budget $\epsilon$ in (\ref{eq:DP_definition1}) is a cumulative privacy budget for these multiple iterations. Sometimes, by fixing the length of the observation sequence to one (a single iteration $k$)  in Definition 2, a  privacy budget  $\epsilon^k$ can be calculated for this particulary iteration $k$ \cite{dwork2006calibrating}. In this case, according to the  sequential composition property of DP \cite{dwork2006calibrating}, $\sum_{k=1}^{\bar{k}}\epsilon^k$ corresponds to $\epsilon$ in our Definition 2 using  observation sequences of length $\bar{k}$ (from $k=1$ to $k=\bar{k}$).}

 $\epsilon$-DP ensures that an adversary having access to all shared messages cannot gain information with a  significant probability of any participating player's cost function. It can  be seen that a smaller $\epsilon$ means a higher level of privacy protection.


\section{Differentially-private consensus tracking}\label{se:algorithm1}
Lying at the core of our distributed GNE seeking algorithm is the consensus-tracking technique, { which enables multiple agents to cooperative track   the average of multiple
time-varying reference signals (i.e., $\frac{\sum_{i=1}^{m}r_i^k}{m}$) while each individual agent $i$ can
only access one reference signal $r_i^k$. (Note that in this section,  $r_i^k$ is regarded as a given signal, and an instantiation will be given in the GNE seeking application in Sec. IV.)}  However,  the conventional consensus-tracking algorithm (see, e.g., \cite{zhu2010discrete}) is sensitive to noise in that it leads to the accumulation and explosion of noise variance   in the presence of persistent information-sharing noise \cite{wang2022gradient}.
Hence,   we first propose a new robust   consensus-tracking algorithm. {Our basic idea is to tailor consensus tracking for $\epsilon$-DP by adding one stepsize parameter $\gamma^k$ and a weakening factor $\chi^k$ (see Algorithm 1), which enable  us to suppress the influence of persistent DP-noise}, and hence,  ensure  both goals of rigorous $\epsilon$-DP and guaranteed  accuracy. 
Following the convention,  we call a participant node an agent here.

\noindent\rule{0.49\textwidth}{0.5pt}
\noindent\textbf{Algorithm 1: Differentially-private consensus tracking with guaranteed convergence accuracy}
\noindent\rule{0.49\textwidth}{0.5pt}
\begin{enumerate}[wide, labelwidth=!, labelindent=0pt]
    \item[] Parameters: Weakening factor $\chi^k>0$ and stepsize $\gamma^k>0$.
    \item[] Every agent $i$'s  reference is $r_i^k$. Every agent $i$ maintains one state variable  $x_i^k$, which is initialized as $x_i^0=r_i^0$.
    \item[] {\bf for  $k=1,2,\ldots$ do}
    \begin{enumerate}
        \item Every agent $j$ adds persistent DP-noise   $\zeta_j^{k}$ 
        to its state
    $x_j^k$,  and then sends the obscured state $x_j^k+\zeta_j^{k}$ to agent
        $i\in\mathbb{N}_j$.
        \item After receiving  $x_j^k+\zeta_j^k$ from all $j\in\mathbb{N}_i$, agent $i$ updates its state  as follows:
        \begin{equation}\label{eq:update_in_Algorithm1}
        \begin{aligned}
              x_i^{k+1} &={(1-\gamma^k)}x_i^k+\chi^k\textstyle\sum_{j\in \mathbb{N}_i} L_{ij}(x_j^k+\zeta_j^k-x_i^k)\\
             &\quad+r_i^{k+1}-{(1-\gamma^k)}r_i^k.
        \end{aligned}
        \end{equation}
        \vspace{-0.2cm}
        \item {\bf end}
    \end{enumerate}
\end{enumerate}
\vspace{-0.2cm} \rule{0.49\textwidth}{0.5pt}
{
\begin{Remark 1}
   Our prior result  in \cite{wang2022aggregate} also discusses enabling DP in a particular consensus-tracking application in aggregative games. However, the approach there leaks information of agent $i$ to a neighboring agent $j$ if agent $j$ is the only  neighbor of agent $i$. This is because at any time instant $k$,   the  noises entering into all agents'  updates  in \cite{wang2022aggregate} always add up to zero. Under such an obfuscation mechanism,  when agent $i$ only has one neighbor (say, agent $j$),  its  information will be leaked to agent $j$:  When $r_i^k$ is   a zero signal,  Theorem 3 in \cite{gao2022algorithm}  shows that the initial state of agent $i$ is inferrable by agent $j$; When   $r_i^k$ is the local gradient of agent $i$  (typical in gradient-tracking
based distributed optimization),  Theorem 3 in \cite{gao2023dynamics}  proves that the final gradient information of agent $i$ is inferrable by its only neighbor agent $j$.  By employing an additional stepsize factor $\gamma^k$ (besides the weakening factor $\chi^k$ for inter-agent coupling), our  Algorithm 1 removes the requirement that the network-level sum of noises entering into  agents’ updates has to be zero, and hence, can avoid the privacy leakage in \cite{wang2022aggregate}.
\end{Remark 1}
}

\begin{Assumption 1}\label{ass:dp-noise}
{For every $i\in[m]$ and every $k$,
the   noises in $\{\zeta_i^k\}$ are  zero-mean independent random variables, and are independent of $\{x_i^0;i\in[m]\}$. The noise variances
 $\mathbb{E}\left[\|\zeta_i^k\|^2 \right]=(\sigma_{i}^k)^2$ satisfy
\begin{equation}\label{eq:condition_assumption1}
\textstyle\sum_{k=0}^\infty (\chi^k)^2\, \max_{i\in[m]}(\sigma_{i}^k)^2 <\infty,
\end{equation}  where $\{\chi^k\}$ is the weakening sequence from Algorithm 1.
The initial random vectors satisfy
$\mathbb{E}\left[\|x_i^0\|^2\right]<\infty$,  $\forall i\in[m]$.}
\end{Assumption 1}

\subsection{Convergence Analysis}

To prove that Algorithm 1 can ensure every $x_i^k$ to track  the average target signal $\bar{r}^k \triangleq\frac{1}{m}\sum_{i=1}^{m} r_i^k$, we first  prove that $\bar{x}^k$ converges almost surely to   $\bar{r}^k$ under the following assumption:
\begin{Assumption 3}\label{as:reference_signal}
For every $i\in[m]$, there exist   some nonnegative sequence $\{\beta^k\}$ and a constant $C$ such that
    \begin{equation}\label{eq:bound_C}
       \|r_i^{k+1}-(1-\gamma^k)r_i^{k}\|\leq \beta^k C
    \end{equation}
  holds,   where $\gamma^k$ is from Algorithm 1 and $\{\beta^k\}$  satisfies $\lim_{k\rightarrow\infty}\frac{\gamma^k}{\beta^k}<\infty$.
\end{Assumption 3}

Note that the condition $\lim_{k\rightarrow\infty}\frac{\gamma^k}{\beta^k}<\infty$  is
necessary since otherwise (\ref{eq:bound_C}) will not hold when $\{r_i^k\}$ is a constant signal.

\begin{Lemma 1}\label{le:bar_x=bar_r}
Under Assumptions \ref{as:L}, \ref{ass:dp-noise},  $\bar{x}^k$ in Algorithm 1 converges {\it a.s.} to $\bar{r}^k$ if the following conditions hold:
\[
 \textstyle\sum_{k=0}^{\infty}\gamma^k= \infty, \quad \sum_{k=0}^{\infty}(\gamma^k)^2<\infty.
\]
\end{Lemma 1}
\begin{proof}

According to the definitions of $\bar{x}^k$ and $\bar{r}^k$, we have the following relationship based on (\ref{eq:update_in_Algorithm1}):
\begin{equation}\label{eq:conservation_average}
\bar{x}^{k+1}=(1-\gamma^k)\bar{x}^k+\chi^k\bar{\zeta}^k+\bar{r}^{k+1}-(1-\gamma^k)\bar{r}^k,
\end{equation}
where $\bar{\zeta}^k \triangleq \frac{\sum_{i=1}^{m} |L_{ii}|\zeta_i^k}{m}$ and we have used the symmetric property of $L$ and the fact   $L_{ii}\triangleq-\sum_{j\in\mathbb{N}_i}L_{ij}$.

The preceding relationship implies
\[
\begin{aligned}
\|\bar{x}^{k+1}-\bar{r}^{k+1}\|^2
&=(1-\gamma^k)^2\|\bar{x}^k-\bar{r}^k\|^2+(\chi^k)^2\|\bar{\zeta}^k\|^2\\
&\qquad+2\left\langle  (1-\gamma^k)(\bar{x}^k-\bar{r}^k),\chi^k\bar{\zeta}^k \right\rangle.
\end{aligned}
\]
Using the assumption on the DP-noise $\zeta_i^k$  in  Assumption~\ref{ass:dp-noise}, taking the conditional expectation, given $\mathcal{F}^k=\{x^0,\,\ldots,x^k\}$,
 we obtain  the following inequality  for all $k\ge0$:
 \[
\begin{aligned}
\mathbb{E}\left[\|\bar{x}^{k+1}-\bar{r}^{k+1}\|^2|\mathcal{F}^k\right]&
 \leq\left(1+(\gamma^k)^2-2\gamma^k\right) \|\bar{x}^k-\bar{r}^k\|^2\\
 &\qquad+\textstyle\frac{\sum_{i=1}^{m}L_{ii}^2(\sigma_i^k)^2(\chi^k)^2}{m},
\end{aligned}
\]
where we have used the relationship $\|\bar{\zeta}^k\|^2\leq \frac{\sum_{i=1}^{m}L_{ii}^2\| \zeta_i^k\|^2}{m}$.

 Under the conditions in the lemma statement, it can be seen that  $\|\bar{x}^{k+1}-\bar{r}^{k+1}\|^2$ satisfies the conditions for $v^k$ in Lemma \ref{Lemma-polyak}, with $o^k=(\gamma^k)^2$, $q^k=2\gamma^k$, and $p^k=\frac{\sum_{i=1}^{m}L_{ii}^2(\sigma_i^k)^2(\chi^k)^2}{m}$. Therefore, we have $\sum_{k=0}^{\infty} \gamma^k\|\bar{x}^{k}-\bar{r}^{k}\|^2<\infty$ and $\|\bar{x}^{k+1}-\bar{r}^{k+1}\|^2$ converging {\it a.s.} to zero, and hence, $\bar{x}^k$ converging {\it a.s.} to $\bar{r}^k$.
 \end{proof}

{
\begin{Remark 1}
   From  the derivation of (\ref{eq:conservation_average}), it can be seen that  $\chi^k$ has to be the same for all agents to make sure that it only attenuates the noise input to $\bar{x}^k$ and does not introduce additional input terms on individual states.
\end{Remark 1}

}

\begin{Lemma 1}\label{Le:rho_2}
Under Assumption \ref{as:L} and two positive sequences $\{\chi^k\}$ and $\{\gamma^k\}$ satisfying $
\sum_{k=0}^\infty (\gamma^k)^2<\infty
$ and $
\sum_{k=0}^\infty (\chi^k)^2<\infty
$, there always exists a ${k_0}\geq 0$ such that
$
\|(1-\gamma^k)(I- \frac{{\bf 1}{\bf 1}^T}{m})+\chi^kL\|< 1-\chi^k |\rho_2|
$
holds for all $k\geq {k_0}$, where $\rho_2$ is the   second largest eigenvalue of $L$.
\end{Lemma 1}
\begin{proof}
Under   Assumption \ref{as:L}  and  the definition
$L_{ii}=-\sum_{j\in\mathbb{N}_i}L_{ij}$, the Gershgorin circle theorem implies  that  all eigenvalues of $L$ are non-positive, with one (and only one) of them being equal to 0. Arrange the eigenvalues of $L$ as $\rho_m\leq \rho_{m-1}\leq \cdots\leq \rho_2<\rho_1=0$. It can be verified that the eigenvalues of $(1-\gamma^k)I+\chi^kL$ are equal to $1-\gamma^k+\chi^k\rho_m\leq 1-\gamma^k+\chi^k\rho_{m-1}\leq \cdots\leq 1-\gamma^k+\chi^k\rho_2 < 1-\gamma^k+\chi^k\rho_1=1-\gamma^k$, and the eigenvalues of $(1-\gamma^k)(I-\frac{{\bf 1}{\bf 1}^T}{m})+\chi^kL=(1-\gamma^k)I+\chi^kL+(1-\gamma^k) \frac{{\bf 1}{\bf 1}^T}{m} $ are given by $\{1-\gamma^k+\chi^k\rho_m,\,1-\gamma^k+\chi^k\rho_{m-1},\, \cdots,\,1-\gamma^k+\chi^k\rho_2,\, 0\}$, with $|\rho_m|\geq|\rho_{m-1}|\geq\cdots\geq |\rho_2|>0$. Hence, the norm of $\|(1-\gamma^k)(I-\frac{{\bf 1}{\bf 1}^T}{m})+\chi^kL\|$ is no larger than $\max\{|1-\gamma^k+\chi^k\rho_m|, |1-\gamma^k+\chi^k\rho_2|\}$. Further taking into account the fact that $\gamma^k$ and $\chi^k$ decay  to zero because they are square summable, we conclude that there always exists a ${k_0}\geq 0$ such that $|1-\gamma^k+\chi^k \rho_m|=1-\gamma^k-\chi^k|\rho_m|<1- \chi^k|\rho_m|$ and $|1-\gamma^k+\chi^k \rho_2|=1-\gamma^k-\chi^k|\rho_2|<1- \chi^k|\rho_2|$ hold   for all $k\geq {k_0}$. Given $|\rho_m|\geq  |\rho_2|$,   we have the stated result.
\end{proof}

\begin{Theorem 1}\label{Th:consensus_tracking}
 Under Assumptions \ref{as:L}, \ref{ass:dp-noise}, \ref{as:reference_signal}, if
   the nonnegative sequences $\{\gamma^k\}$  and $\{\chi^k\}$ in Algorithm 1 and the nonnegative sequence $\{\beta^k\}$ in Assumption \ref{as:reference_signal} satisfy
   \vspace{-0.2cm}
    \begin{equation}\label{eq:condtions_chi}
    \sum_{k=0}^{\infty}\gamma^k=\infty,\, \sum_{k=0}^{\infty}\chi^k=\infty,\, \sum_{k=0}^{\infty}(\chi^k)^2<\infty,\, \sum_{k=0}^{\infty}\frac{(\beta^k)^2}{\chi^k}<\infty,
    \vspace{-0.3cm}
    \end{equation}
   then, the following results hold almost surely:
   \vspace{-0.1cm}
     \begin{enumerate}
    \item  every  $x_i^k$ in Algorithm 1 converges to $\bar{r}^k=\frac{\sum_{i=1}^{m}r_i^k}{m}$;
    \item    $\sum_{k=0}^{\infty}\chi^k\sum_{i=1}^{m}\|x_i^k-\bar{x}^k\|^2<\infty$;
    \item   $\sum_{k=0}^{\infty}\beta^k \sum_{i=1}^{m}\|x_i^k-\bar{x}^k\| <\infty$.
  \end{enumerate}
\end{Theorem 1}

\begin{proof}
For the convenience of analysis, we write the iterates of  $x_i^k$ on per-coordinate expressions. More specifically, for all
$\ell=1,\ldots,d,$ and $k\ge0$, we define
$
x^k(\ell)=\left[[x_1^k]_\ell,\ldots,[x_m^k]_\ell\right]^T
$
where $[x_i^k]_\ell$ represents the $\ell$th element of the vector $x_i^k$. Similarly, we  define
$r^k(\ell)=\left[[r_1^k]_\ell,\ldots,[r_m^k]_\ell\right]^T$  and $\zeta^k(\ell) =\left[[\zeta_1^k]_\ell,\ldots,[\zeta_m^k]_\ell\right]^T$. In
this per-coordinate view, (\ref{eq:update_in_Algorithm1})  has the following form  for all $\ell=1,\ldots,d,$
and $k\ge0$:
\begin{equation}\label{eq:update_percord}
\begin{aligned}
x^{k+1}(\ell)=&(1-\gamma^k)x^k(\ell)+\chi^k L x^k(\ell)+\chi^kL^0 \zeta^k(\ell)  \\
&\quad+r^{k+1}(\ell)-(1-\gamma^k)r^k(\ell),
\end{aligned}
\end{equation}
where $L^0$ is obtained by replacing all diagonal entries of $L$ with zero. 

The dynamics of   $\bar{x}^k$ is  (noting ${\bf 1}^TL=0$ from Assumption \ref{as:L})
%
\begin{equation}\label{eq:bar_x2}
\begin{aligned}
 [\bar{x}^{k+1}]_\ell=& \frac{{\bf 1}^T}{m}\left((1-\gamma^k)x^k(\ell)+ \chi^kL^0 \zeta^k(\ell)\right.\\
 &\qquad\left.+  r^{k+1}(\ell)-(1-\gamma^k) r^k(\ell) \right).
\end{aligned}
\end{equation}

 Combining  (\ref{eq:update_percord})  and (\ref{eq:bar_x2}) yields (with $\Pi\triangleq I-\frac{{\bf 1}{\bf 1}^T}{m}$)
\begin{equation}\label{eq:v_k+1}
\begin{aligned}
&x^{k+1}(\ell)-{\bf 1}[\bar{x}^{k+1}]_{\ell}= \left((1-\gamma^k)\Pi+\chi^kL\right)x^k(\ell)\\
&\quad +\chi^k \Pi L^0\zeta^k(\ell) +\Pi\left(r^{k+1}(\ell)
- (1-\gamma^k)r^k(\ell)\right).
\end{aligned}
\end{equation}

To simplify  notations, we define
$
W^k\triangleq (1-\gamma^k)\Pi+\chi^kL=(1-\gamma^k)(I-\frac{{\bf 1}{\bf 1}^T}{m})+\chi^kL.
$
Assumption \ref{as:L} ensures $ W^k {\bf 1}=0$, and hence $ W^k {\bf 1}[\bar{x}^k]_\ell=0$  for any $1\leq \ell\leq d$. Subtracting $ W^k {\bf 1}[\bar{x}^k]_\ell=0$  from the right hand side of (\ref{eq:v_k+1}) yields
\begin{equation}\label{eq:v_k+1_3}
\begin{aligned}
x^{k+1}(\ell)-{\bf 1}[\bar{x}^{k+1}]_{\ell}=&W^k\left(x^k(\ell)-{\bf 1}[\bar{x}^k]_\ell\right)+\chi^k \Pi L^0\zeta^k(\ell)\\
& +\Pi \left(r^{k+1}(\ell)
- (1-\gamma^k)r^k(\ell)\right),
\end{aligned}
\end{equation}
which, further implies (noting $\|\Pi\|=1$)
\begin{equation}\label{eq:v_k+1_norm}
\begin{aligned}
&\|x^{k+1}(\ell)-{\bf 1}[\bar{x}^{k+1}]_{\ell} \|^2\\
&\leq\left\|W^k (x^k(\ell)-{\bf 1}[\bar{x}^k]_\ell )+\Pi  (r^{k+1}(\ell)
- (1-\gamma^k)r^k(\ell) ) \right\|^2\\
& +2 \big\langle W^k (x^k(\ell)-{\bf 1}[\bar{x}^k]_\ell )+\Pi  (r^{k+1}(\ell)
-  (1-\gamma^k)r^k(\ell) ), \\
&  \qquad \chi^k \Pi L^0\zeta^k(\ell)  \big\rangle+(\chi^k)^2\| L^0\|^2{\|}\zeta^k(\ell) \|^2.
\end{aligned}
\end{equation}

 Using   Assumption~\ref{ass:dp-noise}, taking the conditional expectation, given $\mathcal{F}^k=\{x^0,\,\ldots,x^k\}$,
 we obtain  the following inequality  for all $k\ge0$:
\begin{equation}\label{eq:Ev_k}
\begin{aligned}
&\mathbb{E}\left[\|x^{k+1}(\ell)-{\bf 1}[\bar{x}^{k+1}]_{\ell} \|^2{|}\mathcal{F}^k\right]\\
&\leq\left(\|W^k \|\| x^k(\ell)-{\bf 1}[\bar{x}^k]_\ell  \|+ \|   r^{k+1}(\ell)
- (1-\gamma^k)r^k(\ell)  \|\right)^2\\
&\quad+(\chi^k)^2\|L^0\|^2 \mathbb{E}\left[\|\zeta^k(\ell) \|^2\right].
\end{aligned}
\end{equation}

For the first term on the right hand side of the preceding inequality, we bound it using the fact that there exists a $k_0\geq 0$ such that $0<\|W^k\|\leq 1-\chi^k |\rho_2|$ holds for all $k\geq k_0$ (see Lemma \ref{Le:rho_2}).  Hence,  equation (\ref{eq:Ev_k}) implies   the following relationship   for all $k\geq k_0$:
\begin{equation}\label{eq:Ev_k2}
\begin{aligned}
&\mathbb{E}\left[\|x^{k+1}(\ell)-{\bf 1}[\bar{x}^{k+1}]_{\ell} \|^2|\mathcal{F}^k\right]\leq \\
&\left((1-\chi^k|\rho_2|)\|x^k(\ell)-{\bf 1}[\bar{x}^k]_\ell \|+\|r^{k+1}(\ell)
- (1-\gamma^k)r^k(\ell) \|\right)^2\\
&\quad+(\chi^k)^2\|L^0\|^2\mathbb{E}\left[\|\zeta^k(\ell) \|^2\right].
\end{aligned}
\end{equation}

Next, we apply  the inequality
$(a+b)^2\le (1+\epsilon) a^2 + (1+\epsilon^{-1})b^2$, which is valid for any scalars $a,b,$ and $\epsilon>0$, to (\ref{eq:Ev_k2}).
More specifically, setting $\epsilon=\frac{\chi^k|\rho_2|}{1-\chi^k|\rho_2|}$ (which leads to $(1+\epsilon)=\frac{1}{1-\chi^k|\rho_2|}$ and $1+\epsilon^{-1}=\frac{1}{\chi^k|\rho_2|}$) yields
\begin{equation}\label{eq:Ev_k4}
\begin{aligned}
&\mathbb{E}\left[\|x^{k+1}(\ell)-{\bf 1}[\bar{x}^{k+1}]_{\ell} \|^2|\mathcal{F}^k\right]\leq (\chi^k)^2\|L^0\|^2\mathbb{E}\left[\|\zeta^k(\ell) \|^2\right]\\
&+ (1-\chi^k|\rho_2|) \|x^k(\ell)-{\bf 1}[\bar{x}^k]_\ell\|^2\\
&+ \frac{1}{\chi^k|\rho_2|} \| r^{k+1}(\ell)
- (1-\gamma^k)r^k(\ell)  \|^2.
\end{aligned}
\end{equation}

Note that the following relations always hold:
$
\sum_{\ell=1}^d \|x^k(\ell) - [\bar x^k]_\ell{\bf
1}\|^2=\sum_{i=1}^m\|x^k_i - \bar x^k \|^2,$ $
\sum_{\ell=1}^d \|r^{k+1}(\ell) - (1-\gamma^k)r^{k}(\ell)\|^2=\sum_{i=1}^m\|r^{k+1}_i - (1-\gamma^k)r^k_i\|^2$, $\sum_{\ell=1}^d
\|\zeta^k(\ell)\|^2=\sum_{i=1}^m\|\zeta^k_i\|^2$.

Hence,  summing (\ref{eq:Ev_k4}) over $\ell=1,\ldots,d$ leads to
\begin{equation}\label{eq:Ev_k5}
\begin{aligned}
&\mathbb{E}\left[\sum_{i=1}^{m}\|x_i^{k+1}- \bar{x}^{k+1} \|^2|\mathcal{F}^k\right]\\
&\leq (\chi^k)^2\|L^0\|^2 \sum_{i=1}^{m}(\sigma_i^k)^2+ (1-\chi^k|\rho_2|) \sum_{i=1}^{m}\|x_i^k- \bar{x}^k\|^2\\
&+ \frac{1}{\chi^k|\rho_2|} \sum_{i=1}^{m}\| r_i^{k+1}
- (1-\gamma^k)r_i^k \|^2  .
\end{aligned}
\end{equation}

 Assumption \ref{as:reference_signal} implies
$
\frac{1}{\chi^k|\rho_2|} \sum_{i=1}^{m}\| r_i^{k+1}
- (1-\gamma^k)r_i^k \|^2\leq  \frac{(\beta^k)^2}{\chi^k|\rho_2|}mC^2.
$
 Submitting the   relationship into (\ref{eq:Ev_k5}) yields
\begin{equation}\label{eq:Ev_k6}
\begin{aligned}
&\mathbb{E}\left[\sum_{i=1}^{m}\|x_i^{k+1}- \bar{x}^{k+1} \|^2|\mathcal{F}^k\right]\leq (\chi^k)^2\|L^0\|^2 \sum_{i=1}^{m}(\sigma_i^k)^2\\ &+(1-\chi^k|\rho_2|) \sum_{i=1}^{m}\|x_i^k- \bar{x}^k\|^2  + \frac{(\beta^k)^2}{\chi^k|\rho_2|}mC^2.
\end{aligned}
\end{equation}

Therefore, under Assumption \ref{ass:dp-noise} and the conditions for $\chi^k$ and $\beta^k$ in (\ref{eq:condtions_chi}), we have that the sequence $\{\sum_{i=1}^{m}\|x_i^k- \bar{x}^k\|^2\}$ satisfies the conditions for $\{v^k\}$ in Lemma \ref{Lemma-polyak}, and hence, converges to zero almost surely. So
$x_i^k$ converges to $\bar{x}^k$ almost surely. Further recalling  $\bar{x}^k$ converging {\it a.s.} to $\bar{r}^k$ in Lemma \ref{le:bar_x=bar_r} yields that  $x_i^k$ converges {\it a.s.} to $\bar{r}^k$.

Moreover, Lemma \ref{Lemma-polyak} also implies  the following relation  {\it a.s.}:
\begin{equation}\label{eq:square}
 \sum_{k=0}^{\infty}\chi^k\sum_{i=1}^{m}\|x_i^k- \bar{x}^k\|^2 <\infty.
\end{equation}

To prove the last statement, we invoke the Cauchy–Schwarz inequality, which ensures
$
 \sum_{k=0}^{\infty} \sqrt{\chi^k\sum_{i=1}^{m}\|x_i^k- \bar{x}^k\|^2}\sqrt{\frac{(\beta^k)^2}{\chi^k}}
 \leq \sqrt{\sum_{k=0}^{\infty} \chi^k\sum_{i=1}^{m}\|x_i^k- \bar{x}^k\|^2}\sqrt{\sum_{k=0}^{\infty} \frac{(\beta^k)^2}{\chi^k}}.
$
Noting that the summand in the left hand side of the preceding inequality is actually $\beta^k\sqrt{\sum_{i=1}^{m}\|x_i^k- \bar{x}^k\|^2}$, and the right hand side of the preceding inequality is less than infinity almost surely under the proven result in (\ref{eq:square}) and the assumption in (\ref{eq:condtions_chi}), we have that
$\sum_{k=0}^{\infty}\beta^k\sqrt{\sum_{i=1}^{m}\|x_i^k- \bar{x}^k\|^2}<\infty$ holds almost surely. Further utilizing the relationship
$\sum_{i=1}^{m}\|x_i^k- \bar{x}^k\|\leq  \sqrt{m\sum_{i=1}^{m}\|x_i^k- \bar{x}^k\|^2}$ yields the stated result.
\end{proof}



\subsection{Differential-privacy Analysis}

To analyze the level of DP protection on individual signals $\{r_i^k\}$, we first {define adjacency  for the consensus-tracking problem.  Representing a consensus-tracking problem by two parameters $(r, L)$, where
   $r\triangleq\left\{\{r_1^k\},\,\cdots, \{r_m^k\}\right\}$ and $ L$ is the interaction weight matrix, we define adjacency between two consensus-tracking problems as follows:

  \begin{Definition 1}\label{de:adjacency_consensus}
Two consensus-tracking problems $\mathcal{P}=(r, L)$ and $\mathcal{P}'=(r', L')$ are adjacent if the following conditions hold:
\begin{itemize}
\item  $L=L'$, i.e.,  the interaction weight matrices are identical;
\item there exists an $i\in[m]$ such that $\{r_i^k\}\neq \{{r_i^k}'\}$  but $\{r_j^k\} = \{{r_j^k}'\}$ for all $j\in[m],\,j\neq i$;
\item  the different sequences  $\{r_i^k\}$ and $\{{r_i^k}'\}$  have similar asymptotic behaviors, i.e., there exists some $C_r>0$ such that
\begin{equation}\label{eq:r_convergence_rate}
   \|r^k_i-{r'_i}^k\|_1\leq C_r\chi^k\beta^k
\end{equation}
  holds for all $k\geq 0$, where $\chi^k$ and $\beta^k$ are from Algorithm 1 and Assumption \ref{as:reference_signal}, respectively.
\end{itemize}
\end{Definition 1}
}

{ \begin{Definition 1}
   ($\epsilon$-differential privacy). For a given $\epsilon>0$, an iterative consensus-tracking algorithm  is $\epsilon$-differentially private if for any two adjacent consensus-tracking problems $\mathcal{P}$ and $\mathcal{P}'$, any set of observation sequences $\mathcal{O}_s\subseteq\mathbb{O}$ (with $\mathbb{O}$ denoting the set of all possible observation sequences), and any initial state ${\vartheta}^0$, the following relationship always holds
    \begin{equation}\label{eq:DP_definition}
        \mathbb{P}[\mathcal{R}_{\mathcal{P},\vartheta^0}\in\mathcal{O}_s]\leq e^\epsilon\mathbb{P}[\mathcal{R}_{\mathcal{P}',\vartheta^0}\in\mathcal{O}_s],
    \end{equation}
    with the probability $\mathbb{P}$  taken over the randomness over iteration processes.
 \end{Definition 1}}

From the theory of $\epsilon$-DP,  we  have to characterize the sensitivity of an algorithm in order to quantify the amount of DP-noise that has to be injected to shared messages to enable a certain level of  $\epsilon$-DP. Similar to  the sensitivity definition of iterative   optimization algorithms in \cite{Huang15}, we define  the sensitivity of a consensus-tracking algorithm as follows: 
\begin{Definition 1}\label{de:sensitivity}
  At each iteration $k$, for any initial state $\vartheta^0$ and any adjacent consensus-tracking problems  $\mathcal{P}$ and $\mathcal{P'}$,  the sensitivity of Algorithm 1 is
  \begin{equation}
  \Delta^k\triangleq \sup\limits_{\mathcal{O}\in\mathbb{O}}\left\{\sup\limits_{\vartheta\in\mathcal{R}_{\mathcal{P},\vartheta^0}^{-1}(\mathcal{O}),\:\vartheta'\in\mathcal{R}_{\mathcal{P}',\vartheta^0}^{-1}(\mathcal{O})}\hspace{-0.3cm}\|\vartheta^{k}-\vartheta'^{k}\|_1\right\}.
  \end{equation}
\end{Definition 1}

{ Based on the sensitivity definition, if we just focus on a specific iteration $k$ and use the Laplace noise mechanism \cite{dwork2006calibrating}, the  privacy budget  for this particular iteration $k$ will be  $\epsilon^k=\frac{\Delta^k}{\nu^k}$ when the Laplace noise is with parameter $\nu^k$. Using the sequential composition property of DP \cite{dwork2006calibrating}, the cumulative privacy budget $\sum_{k=1}^{\bar{k}}\epsilon^k$ for iterations from $k=1$ to $k=\bar{k}$ can also be obtained:}
\begin{Lemma 1}\label{Le:Laplacian}
In Algorithm 1, at each iteration $k$, if each agent's DP-noise vector $\zeta_i^k\in\mathbb{R}^d$  consists of $d$ independent Laplace noises with  parameter $\nu^k$   such that $\sum_{k=1}^{\bar{k}}\frac{\Delta^k}{\nu^k}\leq \bar\epsilon$, then Algorithm 1 is {$\bar\epsilon$}-differentially private with the cumulative privacy budget from iteration   $k=1$ to $k=\bar{k}$ less than $\bar\epsilon$.
\end{Lemma 1}
\begin{proof}
The lemma   follows the same line of reasoning of Lemma 2 in  \cite{Huang15} (also see Theorem 3 in \cite{ye2021differentially}). { It can be intuitively understood as follows: At any single iteration $k$, since  the algorithm sensitivity is given by $\Delta^k$,  Laplace noise with parameter $\nu^k$ leads to  DP protection for this single iteration with a single-iteration privacy budget $\frac{\Delta^k}{\nu^k}$ (note that a smaller privacy budget means a stronger privacy protection). Given that the injected noises  are independent across  different iterations, according to the sequential composition property of DP \cite{dwork2006calibrating}, the cumulative privacy budget can be obtained by adding single-iteration privacy budgets in individual iterations. Therefore, for iterations from $k=1$ to $k=\bar{k}$, the cumulative privacy budget is less than $\bar\epsilon$ when $\sum_{k=1}^{\bar{k}}\frac{\Delta^k}{\nu^k}\leq \bar\epsilon$ holds. }
\end{proof}

\begin{Theorem 1}\label{th:DP_Algorithm1}
{Under  the conditions of Theorem \ref{Th:consensus_tracking}}, if all elements of $\zeta_i^k$ are drawn independently from  Laplace distribution ${\rm Lap}(\nu^k)$ with $(\sigma_i^k)^2=2(\nu^k)^2$ satisfying Assumption \ref{ass:dp-noise},  then all agents
will converge {\it a.s.} to the average tracking target $\bar{r}^k$. Moreover,
\begin{enumerate}
\item For any finite number of iterations $\bar{k}$, Algorithm 1 is  $\epsilon$-differentially private with the cumulative privacy budget bounded by $\epsilon\leq \sum_{k=1}^{\bar{k}}\frac{C_r\varsigma^k}{\nu^k}$   where $\varsigma^k\triangleq \sum_{p=1}^{k-1} \Pi_{q=p}^{k-1}(1-\gamma^q-\bar{L}\beta^{q}) \Lambda^{p-1} +\Lambda^{k-1}$, $\bar{L}\triangleq\min_i\{|L_{ii}|\}$, $\Lambda^k \triangleq \beta^{k+1}\chi^{k+1} +  (1-\gamma^k)\beta^{k}\chi^{k}$, and $C_r$ is given in (\ref{eq:r_convergence_rate});
\item  The cumulative privacy budget is  finite for $\bar{k}\rightarrow\infty$  when the sequence  $\{\frac{\beta^k}{\nu^k}\}$ is summable.
\end{enumerate}
\end{Theorem 1}
\begin{proof}
Since the Laplace noise satisfies Assumption \ref{ass:dp-noise} {and we assume that all conditions of Theorem  \ref{Th:consensus_tracking} hold}, the convergence result  follows directly from Theorem 1.

To prove the two statements on the strength of $\epsilon$-DP, we first analyze the sensitivity of the algorithm.
 Given two adjacent  consensus-tracking problems $\mathcal{P}$ and $\mathcal{P'}$, for any given fixed observation $\mathcal{O}$ and initial state $\vartheta^0= x^0$,   the sensitivity depends on $\|x^{k}-x'^{k}\|_1$ according to Definition \ref{de:sensitivity}. Since in $\mathcal{P}$ and $\mathcal{P'}$, only one reference signal  is different, we  represent this different reference signal as the  $i$th one, i.e., $r_i$ in $\mathcal{P}$ and $r'_i$ in $\mathcal{P}'$, without loss of generality.

Because the initial conditions, reference signals, and observations of $\mathcal{P}$ and $\mathcal{P'}$  are identical for $j\neq i$, we have $x_j^k={x'_j}^k$ for all $j\neq i$ and $k$. Therefore, $\|x^{k}-x'^{k}\|_1$ is always equal to $\|x_i^{k}-{x'_i}^{k}\|_1$.

   Algorithm 1   implies
 \[
 \begin{aligned}
x_i^{k+1}-{x'_i}^{k+1}=&(1-\gamma^k-|L_{ii}|\chi^k)(x_i^k-{x'_i}^k)\\
 &+ (r_i^{k+1}-{r'_i}^{k+1})-(1-\gamma^k) (r_i^k-{r'_i}^k).
 \end{aligned}
 \]
 Note that we have  used the  fact that the observations $x_j^k+\zeta_j^k$ and ${x'_j}^k+{\zeta'_j}^k$ are the same.

 Hence, using the third condition in Definition \ref{de:adjacency}, the sensitivity $\Delta^k$ satisfies
 \begin{equation}\label{eq:sensitivity_iterationi}
 \begin{aligned}
 &\Delta^{k+1}\\
 &\leq (1-\gamma^k-|L_{ii}|\chi^k)\Delta^{k}+ C_r\beta^{k+1}\chi^{k+1} + C_r(1-\gamma^k)\beta^{k}\chi^{k}\\
 &\leq (1-\gamma^k-\bar{L}\chi^k)\Delta^{k}+ C_r\Lambda^k,
 \end{aligned}
 \end{equation}
 where  $\bar{L}$ and $\Lambda^k$ are defined in the theorem statement. Then, by iteration, we can arrive at the first privacy statement   using Lemma \ref{Le:Laplacian} (noting $\Delta^0=0$ as $\mathcal{P}$ and $\mathcal{P}'$ have identical initial conditions).

 For the infinity-time-horizon result in the second statement, we exploit Lemma \ref{le:chung} and the third condition in Definition \ref{de:adjacency_consensus}.  More specifically, from (\ref{eq:sensitivity_iterationi}), according to Lemma \ref{le:chung},  we can always find some $\bar{C}$ such that $\Delta^k\leq  \bar{C} \beta^k $ holds (note that $\gamma^k$ decays faster than $\chi^k$).
Using Lemma \ref{Le:Laplacian}, we can easily obtain $\epsilon\leq \sum_{k=1}^{\bar{k}}\frac{\bar{C}\beta^k}{ \nu^k}$. Hence, $\epsilon$ will  be finite even when $\bar{k}$ tends to infinity if  the sequence $\{\frac{\beta^k}{\nu^k}\}$ is summable.
\end{proof}

Note that in consensus-tracking applications such as constraint-free Nash equilibrium seeking, \cite{ye2021differentially} achieves  $\epsilon$-DP by enforcing the tracking reference   to be summable (geometrically-decreasing, more specifically), which, however, also makes it impossible to ensure accurate convergence to the  desired equilibrium. In our approach, by allowing the tracking reference $r_i^{k+1}-(1-\gamma^k)r_i^k$  to be non-summable (since Assumption \ref{as:reference_signal}   allows  $\sum_{k=0}^{\infty}\beta^k=\infty$), we achieve both accurate convergence and finite cumulative privacy budget.  To our knowledge, this is the first consensus-tracking algorithm that can achieve both {\it a.s.} convergence to a tracking target  and rigorous $\epsilon$-DP, even on the infinite time horizon.

\begin{Remark 1}
  To ensure that the cumulative privacy budget $\epsilon=\sum_{k=1}^{\infty}\frac{\bar{C}\beta^k}{\nu^k}$ is bounded, we  employ   Laplace noise with parameter $\nu^k$  increasing with time (since we require the sequence $\{\frac{\beta^k}{\nu^k}\}$ to be summable while the sequence $\{\beta^k\}$ is non-summable).  Because { when sending signals, every agent $i$ always uses $x_i^k$, the amplitude of which is time-invariant}, an increasing $\nu^k$  makes the  { noise-to-signal amplitude ratio} increase  with time. However, since it is $\chi^k{\rm Lap}(\nu^k)$ that is actually  fed into the algorithm, and  the increase  in the noise level  $\nu^k$ is outweighed by the decrease of $\chi^k$ (see  Assumption \ref{ass:dp-noise}), the actual noise fed into the algorithm  still decays with time, which explains why  Algorithm 1 can ensure every agent's  accurate convergence. Moreover, according to Theorem \ref{Th:consensus_tracking}, the convergence will not be affected if we  scale  $\nu^k$ by any constant  $\frac{1}{\epsilon}>0$  to achieve any desired level of $\epsilon$-DP, as long as  $\nu^k$ (with associated variance $(\sigma_i^k)^2=2(\nu^k)^2$) satisfies Assumption \ref{ass:dp-noise}. 
\end{Remark 1}

\section{Differentially-private   GNE seeking}\label{se:algorithm_2}

Based on our differentially-private consensus-tracking algorithm, we propose a differentially-private GNE seeking algorithm that can ensure both rigorous $\epsilon$-DP and provable convergence to the GNE, which is summarized in Algorithm 2. Inspired by the distributed GNE seeking algorithm in \cite{belgioioso2020distributed},  we  let each player maintain three estimates of the global information that are not locally available. More specifically, $\sigma_i^k$ is used to track the average $\bar{x}^k$ of the decision variables $x_i^k$ of all   players, $z_i^k$ is used to track the average $\bar{\lambda}^k$ of the dual variables $\lambda_i^k$, and  $y_i^k$ is used to track the average $\bar{d}^k$ of the violation of the coupling constraint, which is defined as
\begin{equation}\label{eq:d_i^k}
d_i^k\triangleq 2C_i\tilde{x}_i^k-C_ix_i^k-c_i
\end{equation}
for player $i$ ({ where $\tilde{x}_i^k$ is an auxiliary variable with update rule given in the first line in (\ref{eq:update_algo2})}).

\noindent\rule{0.49\textwidth}{0.5pt}
\noindent\textbf{Algorithm 2: Differentially-private GNE seeking algorithm
with guaranteed convergence accuracy}
\noindent\rule{0.49\textwidth}{0.5pt}
\begin{enumerate}[wide, labelwidth=!, labelindent=0pt]
    \item[] Parameters: Weakening factor $\chi^k>0$ and stepsizes $\alpha^k, \gamma^k>0$.
    \item[] Every player $i$ maintains a decision variable  $x_i^k$ ({ with  values at $k=-1$ and $k=0$, i.e.,     $x_i^{-1}$ and $x_i^0$}, selected randomly in $\Omega_i\subseteq\mathbb{R}^d$) and  an auxiliary variable $\tilde{x}_i^k$ ({ with  value at $k=0$, i.e., $\tilde{x}_i^0$, selected randomly in $\Omega_i\subseteq\mathbb{R}^d$)}. Player $i$ also maintains a dual variable $\lambda_i^k$, which is initialized randomly in $\mathbb{R}_{+}^n$. Every player $i$  maintains local estimates (represented by $\sigma_i^k$, $z_i^k$, and $y_i^k$) of global information $\bar{x}^k$, $\bar{\lambda}^k$, and $\bar{d}^k$. These estimates are initialized as $\sigma_i^0=x_i^0$, $z_i^0=\lambda_i^0$, and $y_i^0=2C_i\tilde{x}_i^{-1}-C_ix_i^{-1}-c_i$, respectively.

    \item[] {\bf for  $k=0,1,\ldots$ do}
    \begin{enumerate}
        \item Every player  $j$  adds persistent DP-noises   $\zeta_j^{k}$, $\xi_j^k$, and ${\upsilon_j^k}$ to $\sigma_j^k$, $y_j^k$, and $z_j^k$, respectively, and then  sends the obscured estimates $\sigma_j^k+\zeta_j^{k}$, $y_j^k+\xi_j^k$, and $z_j^k+\upsilon_j^k$ to player
        $i\in\mathbb{N}_j$.
        \item After receiving  $\sigma_j^k+\zeta_j^k$, $y_j^k+\xi_j^k$, and $z_j^k+\upsilon_j^k$  from all $j\in\mathbb{N}_i$, player $i$ updates its decision variable and estimates  as follows:
        \begin{equation}\label{eq:update_algo2}
        \begin{aligned}
        \tilde{x}_i^{k}&=\Pi_{\Omega_i}\left[x_i^k-\alpha^k (F_i(x_i^k,\sigma_i^k)+C_i^Tz_i^k)\right],\\
        y_i^{k+1}&= {(1-\gamma^k)}y_i^k+\chi^k\textstyle\sum_{j\in \mathbb{N}_i}L_{ij}(y_j^k+\xi_j^k-y_i^k) \\
        &\qquad +d_i^k-{ (1-\gamma^k)}d_i^{k-1},\\
        \tilde\lambda_i^k&=\Pi_{\mathbb{R}^n_{+}}\left[ \lambda_i^k+{ \alpha}^k(y_i^{k+1}-\lambda_i^k+z_i^k) \right],\\
        x_i^{k+1}&=x_i^k+\gamma^k(\tilde{x}_i^k-x_i^k),\\
        \lambda_i^{k+1}&=\lambda_i^k+\gamma^k(\tilde{\lambda}_i^k-\lambda_i^k),\\
        \sigma_i^{k+1}&={ (1-\gamma^k)}\sigma_i^k+\chi^k\textstyle\sum_{j\in \mathbb{N}_i}L_{ij}(\sigma_j^k+\zeta_j^k-\sigma_i^k)\\
        &\quad+x_i^{k+1}-{ (1-\gamma^k)}x_i^k,\\
        z_i^{k+1}&={ (1-\gamma^k)}z_i^k+\chi^k\textstyle\sum_{j\in \mathbb{N}_i}L_{ij}(z_j^k+\upsilon_j^k-z_i^k)\\
        &\quad+\lambda_i^{k+1}-{ (1-\gamma^k)}\lambda_i^k.
        \end{aligned}
        \end{equation}
        where $\Pi_{\dag}[\cdot]$ denotes the Euclidean  projection of a vector onto a set $\dag$.
                \item {\bf end}
    \end{enumerate}
\end{enumerate}
\vspace{-0.1cm} \rule{0.49\textwidth}{0.5pt}


{
\begin{Remark 1}
In many applications, constraint parameters contain sensitive information (e.g., local power generation in smart grid \cite{chen2022privacy}) and should be kept private. This motivates us to mask all   transmitted variables   with noises, including  dual variables  corresponding to constraints. In fact, in many cases,   the iteration trajectory of dual variables still bears information of the primal variable, which makes it necessary to mask these variables. This is particularly the case in primal-dual subgradient  methods like \cite{tjell2019privacy} and distributed dual  subgradient methods like \cite{han2021privacy} where disclosing dual updates in two consecutive iterations allows an observer to directly calculate some function value of the primal variable.
\end{Remark 1}
}

 $\{\chi^k\}$, which diminishes with time, is used to suppress the influence of  DP-noises $\zeta_i^k$, $\xi_i^k$, and $\upsilon_i^k$ on the convergence point.
The stepsizes $\{\alpha^k\},   \{\gamma^k\}$ and attenuation sequence $\{\chi^k\}$
have to be designed appropriately to guarantee the accurate convergence of the iterate vector $x^k\triangleq{\rm col}(x_1^k,\cdots,x_m^k)$ to the GNE $x^{\ast}\triangleq {\rm col}(x_1^\ast,\cdots,x_m^\ast)$.
The  DP noise sequences $\{\zeta_i^k\},\, \{\xi_i^k\},\, \{\upsilon_i^k\},\, i\in[m]$ satisfy the assumption below:

\begin{Assumption 1}\label{ass:dp-noises-game}
In Algorithm 2, for every $i\in[m]$, the noise sequences $\{\zeta_i^k\}$,
$\{\xi_i^k\}$, and $\{\upsilon_i^k\}$ are zero-mean independent random variables, and are independent of $\{\sigma_i^0,z_i^0,y_i^0;i\in[m]\}$.
Also, for every $k$, the noise collection $\{\zeta_j^k,\xi_j^k,\upsilon_j^k; j\in[m]\}$ is independent. The noise variances
$(\sigma_{\zeta,i}^k)^2=\mathbb{E}\left[\|\zeta_i^k\|^2\right]$, $(\sigma_{\xi,i}^k)^2=\mathbb{E}\left[\|\xi_i^k\|^2\right]$, and $(\sigma_{\upsilon,i}^k)^2=\mathbb{E}\left[\|\upsilon_i^k\|^2\right]$
satisfy
\vspace{-0.2cm}
\begin{equation}\label{eq:condition_assumption5}
\begin{aligned}
\hspace{-0.3cm}\textstyle\sum_{k=0}^\infty(\chi^k)^2&\max_{i\in[m]}\left\{\max\left\{(\sigma_{\zeta,i}^k)^2,(\sigma_{\xi,i}^k)^2,(\sigma_{\upsilon,i}^k)^2\right\}\right\}<\infty.
\end{aligned}
\end{equation}
The initial random vectors satisfy
$\max\{\mathbb{E}\left[\|x_i^0\|^2\right],\, \mathbb{E}\left[\|z_i^0\|^2\right],\,\mathbb{E}\left[\|y_i^0\|^2\right]\}<\infty$,  $\forall i\in[m]$.
\end{Assumption 1}

\section{Convergence Analysis of Algorithm 2}\label{se:convergence_algo2}

Our  convergence analysis  leverages  stochastically-perturbed  nonstationary fixed-point iteration processes. Recently, the authors in \cite{combettes2015stochastic} show that stochastically-perturbed nonstationary fixed-point iteration processes with nonexpansive operators can still converge to a fixed point if the iteration stepsize sequence $\{\gamma^k\}$ is lower bounded away from $0$, i.e., $\lim\inf_{k\rightarrow\infty} {\gamma^k}>0$. In our case, since the DP-noises are persistent  to ensure a strong privacy, the stepsize has to decay to zero to ensure a finite privacy budget. Hence, we extend the result in \cite{combettes2015stochastic} to the case where the iteration stepsize  decays to zero. Interestingly, we prove that {\it a.s.} convergence to a fixed point can still be achieved in this case.
It is worth noting that our results generalize the Krasnosel'ski\u{i}-Mann iteration process in \cite{bauschke2011convex} which addresses a time-invariant operator, and the inexact Krasnosel'ski\u{i}-Mann iteration process in \cite{belgioioso2020distributed,combettes2001quasi} which addresses deterministic errors.

We first prove that a stochastically-perturbed nonstationary fixed-point iteration process  can still ensure {\it a.s.} convergence to a fixed point, even when the   stepsize decays to zero:

\begin{Lemma 1}\label{le:perburbed_fixed_point}
  {  If there exists some $k_0\geq 0$ such that }for every $k\geq k_0$,   $R_k:\mathbb{R}^{p}\rightarrow \mathbb{R}^{p}$ are nonexpansive operators and ${\bf{F}}\triangleq\bigcap_{k\geq k_0}{\rm Fix}(R_k)\neq \emptyset$, then, under iteration
  \begin{equation}\label{eq:iterate}
    \omega^{k+1}=\omega^k+\gamma^k\left(R_k(\omega^k)+e^k-\omega^k\right),
  \end{equation}
 $\omega^k$ converges {\it a.s.} to some $\omega^\ast\in \bf{F}$ when the following conditions are satisfied:
  \begin{enumerate}
    \item The nonnegative sequence $\{\gamma^k\}$ satisfies $\sum_{k=0}^{k}\gamma^k(1-\gamma^k)=+\infty$;
    \item The error sequence $\{e^k\}$ satisfies $\sum_{k=0}^{\infty}\gamma^k\sqrt{\mathbb{E}\left[\|e^k\|^2|\mathcal{F}^k\right]}<\infty$ almost surely, where $\mathcal{F}^k=\{\omega^\ell; 0\le \ell\le k\}$;
    \item For every $\{\omega^k\}$, $\lim_{k\rightarrow\infty}\|R_k(\omega^k)-\omega^k\|$ exists almost surely;
    \item The cluster points of $\{\omega^k\}$ belong to $\bf F$ almost surely.
  \end{enumerate}
\end{Lemma 1}
\begin{proof}
  This lemma is an adaptation of Corollary 2.7 and Theorem 3.2 of \cite{combettes2015stochastic}. More specifically, we use the {\it a.s.} convergence of $\lim_{k\rightarrow\infty}\|R_k(x^k)-x^k\|$ in  condition  3) to replace the requirement of $\lim\inf_{k\rightarrow\infty} {\gamma^k}>0$ in Theorem 3.2 of \cite{combettes2015stochastic}, which serves the same purpose in the derivation. {  It is  worth noting that since the results in Corollary 2.7 and Theorem 3.2 of \cite{combettes2015stochastic} are asymptotic, they remain
valid when the starting index is shifted from $k=0$ to $k=k_0$, for an arbitrary $k_0\geq 0$. }
\end{proof}
{
\begin{Remark 1}
  Since we do not impose any extra conditions on $R_k$ except  being nonexpansive, we introduce  Condition 3) and  Condition 4) to ensure convergence behaviors. The two conditions can be proven to be satisfied for the particular $R_k$ corresponding to our proposed algorithms.
\end{Remark 1}
}

\begin{Remark 1}
 Different from  \cite{combettes2015stochastic} which requires the stepsize  in nonstationary fixed-point iteration processes to be bounded away from 0, Lemma \ref{le:perburbed_fixed_point} allows the stepsize to decay to zero. This difference is significant in that allowing the stepsize to decay to zero is key to ensuring the privacy budget to be bounded, as detailed later in Theorem \ref{th:DP_Algorithm2}.
\end{Remark 1}

  In the remainder of the convergence analysis, we show that Algorithm 2 is indeed a stochastically-perturbed version of a nonstationary fixed-point iteration process. Hence, we can leverage Lemma \ref{le:perburbed_fixed_point} to establish the {\it a.s.} convergence of Algorithm 2 under persistent DP-noises. The main challenges lie in 1)  finding a sequence  of deterministic nonexpansive operators $R_k(\cdot)$ that  have convergent properties as required in the third and fourth conditions in Lemma \ref{le:perburbed_fixed_point}, and 2) proving that Algorithm 2 is a stochastically-perturbed version of $R_k(\cdot)$ with the amplitude of perturbation $e^k$ satisfying the {\it a.s.} summability condition in condition 2) of Lemma \ref{le:perburbed_fixed_point}. We address the two challenges in Sec. \ref{se:non-perburbed} and Sec. \ref{se:perturbation_summable}, respectively.

\subsection{Finding a non-perturbed  fixed-point iteration process}\label{se:non-perburbed}

 In this subsection, we show that when the estimates $\sigma_i^k$, $y_i^k$, and $z_i^k$ in Algorithm 2 are replaced with their target signals that they are designed to track, i.e., $\bar{x}^k$, $\bar{d}^k$, and $\bar{\lambda}^k$, then the iterations in Algorithm 2 reduce  to a nonstationary fixed-point iteration process with involved  operators being nonexpansive and satisfying the   conditions 3) and 4) in Lemma \ref{le:perburbed_fixed_point}, if  $\{\alpha^k\}$ is set as a nonnegative sequence  satisfying $\sum_{k=0}^{\infty}\alpha^k=\infty$ and $\sum_{k=0}^{\infty}(\alpha^k)^2<\infty$. Note that replacing $\sigma_i^k$, $y_i^k$, and $z_i^k$ with $\bar{x}^k$, $\bar{d}^k$, and $\bar{\lambda}^k$ corresponds to the full-decision information setting where a central coordinator  has direct access to all ${x}_i^k$, $d_i^k$, and $\lambda_i^k$ for $i\in[m]$, and hence, can compute  and disperse  the respective averages to all players.   The reduced algorithm is summarized in Algorithm 3:

   \noindent\rule{0.49\textwidth}{0.5pt}
\noindent\textbf{Algorithm 3:  GNE seeking  with full-decision information}

\noindent\rule{0.49\textwidth}{0.5pt}
\begin{enumerate}[wide, labelwidth=!, labelindent=0pt]
    \item[] Parameters: Stepsizes $\alpha^k>0$   and $\gamma^k>0$.
    \item[] Every player $i$ maintains one decision variable  $x_i^k$, which is initialized   randomly  in $\Omega_i\subseteq\mathbb{R}^d$. Every player $i$ also maintains  a dual variable $\lambda_i^k$, which is initialized randomly in $\mathbb{R}_{+}^n$.
    \item[] {\bf for  $k=0,1,\ldots$ do}
    \begin{enumerate}

        \item Every player $i$ calculates
             \begin{equation}\label{eq:update_in_Algorithm3}
        \begin{aligned}
        \tilde{x}_i^{k}&=\Pi_{\Omega_i}\left[x_i^k-\alpha^k (F_i(x_i^k,\bar{x}^k)+C_i^T\bar{\lambda}^k)\right],\\
        d_i^k&= 2C_i\tilde{x}_i^k-C_ix_i^k-c_i,\\
             \tilde\lambda_i^{k}&=\Pi_{\mathbb{R}^n_{+}}\left[ \lambda_i^k+{ \alpha}^k(\bar{d}_i^k-\lambda_i^k+\bar{\lambda}^k) \right].
        \end{aligned}
        \end{equation}

        \item Player $i$ updates its  variables   as follows:
            \begin{equation}\label{KM_process}
                  x_i^{k+1} =x_i^k+\gamma^k(\tilde{x}_i^k-x_i^k), \:
                  \lambda_i^{k+1} =\lambda_i^k+\gamma^k(\tilde{\lambda}_i^k-{ \lambda_i^k}).
            \end{equation}
                \item {\bf end}
    \end{enumerate}
\end{enumerate}
\vspace{-0.1cm} \rule{0.49\textwidth}{0.5pt}

 Following \cite{belgioioso2020distributed}, we have the following results:
%
\begin{Proposition 1}\label{Le:bounded_iterates}
In both Algorithm 2 and Algorithm 3, $x_i^k$, $\tilde{x}_i^k$, $\tilde{\lambda}_i^k$, $\lambda_i^k$  can always be made bounded.
\end{Proposition 1}
\begin{proof}
  One can obtain that $\tilde{x}_i^k$ and $x_i^k$ are always within $\Omega_i$ since $\tilde{x}_i^k$ is always obtained by projection onto  $\Omega_i$, and the update $x_i^k=x_i^{k-1}+\gamma^k(\tilde{x}_i^{k-1}-x_i^{k-1})$ amounts to a convex combination of elements in the convex set $\Omega_i$. Since $\Omega_i$ is compact, we have that $\tilde{x}_i^k$ and $x_i^k$ are bounded. The boundedness of $\tilde{\lambda}_i^k$ can be easily enforced   in a distributed manner by a superset based technique (commonly used in distributed optimization \cite{chang2014distributed} and distributed GNE seeking \cite{belgioioso2020distributed}), which {  enables us to   replace the projection set $\mathbb{R}^n_+$ in the update rule of $\tilde{\lambda}_i^k$ with a convex compact set. It is worth noting that this replacement does not affect the convergence analysis or the DP-design.}  Under the update rule $\lambda_i^k=\lambda_i^{k-1}+\gamma^k(\tilde{\lambda}_i^{k-1}-\lambda_i^{k-1})$, a bounded $\tilde\lambda_i^k$ implies a bounded $\lambda_i^k$.
\end{proof}

\begin{Lemma 1}\label{le:R_k_algo2}
  Algorithm 3 corresponds to a  nonstationary fixed-point iteration process
    \begin{equation}\label{eq:iterate_algo3}
    \omega^{k+1}=\omega^k+\gamma^k\left(R_k(\omega^k)-\omega^k\right),
  \end{equation}
   where
   $\omega={\rm col}(x^k,\lambda^k)$, and the operators $R_k$ are given by
   \begin{equation}\label{eq:R_k}
   R_k=\left({\rm Id}+(\Phi^k)^{-1}T_2\right)^{-1}\circ\left({\rm Id}-(\Phi^k)^{-1}T_1\right),
   \end{equation}
   with $T_1$ and $T_2$ given in (\ref{eq:T_1+T_2}) and
    \begin{equation}\label{Phi^k}
    \Phi^k=\left[\begin{array}{cc}(\alpha^k)^{-1}I_{md}&-\frac{C_f^T}{m}\\-\frac{C_f}{m}& (\alpha^k)^{-1}I_{mn}\end{array}\right].
 \end{equation}  Moreover, when the non-increasing sequence $\{\alpha^k\}$ satisfies $\sum_{k=0}^{\infty}\alpha^k=\infty$,  $\sum_{k=0}^{\infty}(\alpha^k)^2<\infty$, and
 \begin{equation}\label{eq:alpha_bound}
   \alpha^k\leq \frac{m}{2\max_{i\in [m]}{\|C_i\|}},
 \end{equation}
 { there exists some $k_0\geq 0$ such that for all $k\geq k_0$, }
  $\{R_k\}$
    are nonexpansive operators and the conditions  3) and 4) in Lemma \ref{le:perburbed_fixed_point} are satisfied.
\end{Lemma 1}

\begin{proof}

Algorithm 3 can be  written in the following compact form:
\begin{equation}\label{eq:alg3_compact}
 \begin{aligned}
 \tilde{x}^k&=\Pi_\Omega\left[x^k-\alpha^k\left(F(x^k,\bar{x}_{\bf 1}^k)+C_d^T\bar{\lambda}_{\bf 1}^k\right)\right],\\
 \tilde{\lambda}^k&=\Pi_{\mathbb{R}^{mn}_{+}}\left[\lambda^k{ +\alpha}^k\left(\bar{d}^k_{\bf 1}-\lambda^k+\bar{\lambda}^k_{\bf 1}\right)\right],\\
 x^{k+1}&=x^k+\gamma^k(\tilde{x}^k-x^k),\\
 \lambda^{k+1}&=\lambda^k+\gamma^k(\tilde{\lambda}^k-\lambda^k),
 \end{aligned}
\end{equation}
where
$\bar{x}_{\bf 1}^k={\bf 1}\otimes\bar{x}^k$, $\bar{\lambda}_{\bf 1}^k={\bf 1}\otimes\bar{\lambda}^k$, $\bar{d}_{\bf 1}^k={\bf 1}\otimes\bar{d}^k$, and $C_d={\rm diag}(C_1,\cdots,C_m)$. Then, following the forward-backward operator splitting approach in \cite{yi2019operator,belgioioso2020distributed}, we can obtain that  Algorithm 3 corresponds to a nonstationary  fixed-point iteration process (\ref{eq:iterate_algo3}).

{ Next, we prove that there exists a $k_0\geq 0$ such that $R_k$s are nonexpansive for all $k\geq k_0$.}
According to the Gershgorin circle theorem, the eigenvalues of $\Phi^k$ are no less than $\min_{i\in[m]}\{\frac{1}{\alpha^k}-\frac{\|C_i\|}{m}\}$. Hence, we have that the eigenvalues of $\Phi^k$ are no less than  $\frac{1}{2\alpha^k}$ according to (\ref{eq:alpha_bound}), implying $\|\Phi^k\|\geq \frac{1}{2\alpha^k}$ and $\|(\Phi^k)^{-1}\|\leq  2\alpha^k$.
{  Using this property of $(\Phi^k)^{-1}$, we proceed to prove that there exists a $k_1\geq 0$ such that for all $k\geq k_1$,
$(\Phi^k)^{-1}T_1$ is  $\frac{\delta}{2\alpha^k}$-cocoercive in the Euclidean norm  for all $0<\delta\leq \min\{1,\frac{\mu}{\tilde{L}^2}\}$, where $\mu$ and $\tilde{L}$ are from Assumption \ref{ass:cocoersive} and Assumption \ref{ass:Lipschitz}, respectively.
  To this end, using the matrix inversion lemma (\cite{higham2002accuracy}, page 258), we   rewrite $(\Phi^k)^{-1}$ explicitly as follows:
  \begin{equation}\label{Phi^-1}
  \begin{aligned}
   \left(\Phi^k\right)^{-1}&= I_{\alpha^k} \left(I -U\left(I_\diamond^{-1}+ I_{\alpha^k}U\right)^{-1}I_{\alpha^k} \right),
   \end{aligned}
 \end{equation}
 where
 $ I_{\alpha^k} \triangleq \alpha^k I_{m(d+n)}$,
$ U $ is a block diagonal matrix $U\triangleq {\rm diag}(-\frac{C_f^T}{m},-\frac{C_f}{m})$, and $I_\diamond$ is an anti-diagonal block matrix $I_\diamond\triangleq  [0, I_{md};I_{mn}, 0]$.

Defining $M^k$ as $U(I_\diamond^{-1}+\alpha^kU)^{-1}$, we can obtain the following relation from (\ref{Phi^-1}):
\begin{equation}\label{eq:Phi_1}
  \left(\Phi^k\right)^{-1}=\alpha^k(I-\alpha^kM^k).
\end{equation}

According to the properties of cocoercive mapping, the relationship in (\ref{eq:Phi_1}) implies that   to prove   $(\Phi^k)^{-1}T_1$ is $\frac{\delta}{2\alpha^k}$-cocoercive, we only need to prove $(I-\alpha^kM^k)T_1$ being $\frac{\delta}{2}$-cocoercive, i.e.,
\begin{equation}\label{eq:co_2}
\begin{aligned}
  &\left\langle (I-\alpha^kM^k)(T_1(\omega_1)-T_1(\omega_2)),\omega_1-\omega_2 \right\rangle \\
  &\geq \frac{\delta}{2}\| (I-\alpha^k M^k)(T_1(\omega_1)-T_1(\omega_2))\|^2
\end{aligned}
\end{equation}
for all $\omega_1,\omega_2\in \Omega\times E$ where $\Omega=\Omega_1\times\cdots\times\Omega_m$ and $E$ denotes the space of  augmented dual variables $\rm{col}(\lambda_1,\cdots,\lambda_m)$.

Since $F(\cdot)$ is $\frac{\mu}{\tilde{L}^2}$-cocoercive in the Euclidean norm, following the same proof of  Lemma 3(ii) in \cite{belgioioso2020distributed}, $T_1$ is $\delta$-cocoercive for any $0<\delta\leq \min\{1,\frac{\mu}{\tilde{L}^2}\}$.
Hence, the left hand side of (\ref{eq:co_2}) satisfies
\begin{equation}\label{eq:co_2_left}
\begin{aligned}
  &\left\langle (I-\alpha^kM^k)(T_1(\omega_1)-T_1(\omega_2)),\omega_1-\omega_2 \right\rangle\\
=&\langle  T_1(\omega_1)-T_1(\omega_2),\omega_1-\omega_2 \rangle\\
  &-\alpha^k\langle  M^k(T_1(\omega_1)-T_1(\omega_2)),\omega_1-\omega_2 \rangle\\
 \geq  &\delta \|T_1(\omega_1)\hspace{-0.03cm}-\hspace{-0.03cm}T_1(\omega_2)\|^2-\hspace{-0.075cm}\alpha^k\langle  M^k(T_1(\omega_1)-T_1(\omega_2)),\omega_1-\omega_2 \rangle\\
 \geq  &\delta \|T_1(\omega_1)\hspace{-0.05cm}-\hspace{-0.05cm}T_1(\omega_2)\|^2\hspace{-0.1cm}-\hspace{-0.08cm}\alpha^k \|M^k\|\|T_1(\omega_1)\hspace{-0.05cm}-\hspace{-0.05cm}T_1(\omega_2)\|\|\omega_1\hspace{-0.05cm}-\hspace{-0.05cm}\omega_2\|.
  \end{aligned}
\end{equation}

The right hand side of (\ref{eq:co_2}) can be verified to satisfy
\begin{equation}\label{eq:co_2_right}
  \begin{aligned}
  &\frac{\delta}{2}\| (I-\alpha^k M^k)(T_1(\omega_1)-T_1(\omega_2))\|^2\\
  =&\frac{\delta}{2}\|T_1(\omega_1)-T_1(\omega_2) \|^2+\frac{\delta(\alpha^k)^2}{2}\|M^k(T_1(\omega_1)-T_1(\omega_2))\|^2\\
  &+ \delta\alpha^k \left\langle T_1(\omega_1)-T_1(\omega_2), M^k(T_1(\omega_1)-T_1(\omega_2))\right\rangle\\
  \leq &\frac{\delta}{2}\big(1+2\alpha^k\|M^k\|+(\alpha^k)^2\|M^k\|^2\big)\|T_1(\omega_1)-T_1(\omega_2)\|^2.
  \end{aligned}
\end{equation}
Therefore, to prove the relation in (\ref{eq:co_2}), we only need to prove that (\ref{eq:co_2_left}) is  no smaller than (\ref{eq:co_2_right}), or (after combining like terms) the following inequality
\begin{equation}\label{eq:co_3}
\begin{aligned}
&\frac{\delta}{2}\big(1-2\alpha^k\|M^k\|-(\alpha^k)^2\|M^k\|^2\big)\|T_1(\omega_1)-T_1(\omega_2)\|^2\geq \\
&\geq \alpha^k \|M^k\|\|T_1(\omega_1)\hspace{-0.05cm}-\hspace{-0.05cm}T_1(\omega_2)\|\|\omega_1\hspace{-0.05cm}-\hspace{-0.05cm}\omega_2\|.
\end{aligned}
\end{equation}
Since (\ref{eq:co_3}) always holds under $\|T_1(\omega_1)-T_1(\omega_2)\|=0$, proving (\ref{eq:co_3}) is equivalent to proving
\begin{equation}\label{eq:co_4}
\begin{aligned}
&\frac{\delta}{2}\big(1-2\alpha^k\|M^k\|-(\alpha^k)^2\|M^k\|^2\big)\|T_1(\omega_1)-T_1(\omega_2)\| \\
&\geq \alpha^k \|M^k\| \|\omega_1\hspace{-0.05cm}-\hspace{-0.05cm}\omega_2\|
\end{aligned}
\end{equation}
under the condition  $\|T_1(\omega_1)-T_1(\omega_2)\|\neq 0$.

Hence, proving $(\Phi^k)^{-1}T_1$ to be $\frac{\delta}{2\alpha^k}$-cocoercive in the Euclidean norm when $k$ is larger than some $k_1$ reduces to proving
\begin{equation}\label{eq:co_5}
\begin{aligned}
&\|T_1(\omega_1)-T_1(\omega_2)\|  \\
&\geq \frac{\alpha^k \|M^k\|}{\frac{\delta}{2}(1-2\alpha^k\|M^k\|-(\alpha^k)^2\|M^k\|^2)} \|\omega_1\hspace{-0.05cm}-\hspace{-0.05cm}\omega_2\|
\end{aligned}
\end{equation}
for all $k\geq k_1$ under the condition  $\|T_1(\omega_1)-T_1(\omega_2)\|\neq 0$ (note that $\|M^k\|$ is always bounded and $\alpha^k$ tends to zero under the Lemma conditions, which ensure $1-2\alpha^k\|M^k\|-(\alpha^k)^2\|M^k\|^2 >0$  when $k_1$ is   sufficiently large).

Recalling the definition of $\omega$, i.e., $\omega_1 \triangleq {\rm col}(x_1, \lambda_1)$ and $\omega_2 \triangleq {\rm col}(x_2, \lambda_2)$, the condition  $\|T_1(\omega_1)-T_1(\omega_2)\|\neq 0$ implies either 1) $x_1\neq x_2$, or 2) $\lambda_1-\lambda_2\notin     E^{||}$ where $  E^{||}$ denotes the consensus subspace of the dual variables $\lambda={\rm col}(\lambda_1,\cdots,\lambda_m)$. Denoting $ E^{\perp}$ as the disagreement subspace of the dual variables $\lambda$, we can decompose  $\lambda_1-\lambda_2$ as $\lambda_1-\lambda_2=(\lambda_1-\lambda_2)_{||}+(\lambda_1-\lambda_2)_\perp$ following the argument in the proof of Lemma 3 in \cite{belgioioso2020distributed}, where $(\lambda_1-\lambda_2)_{||}\in  E^{||}$ and $(\lambda_1-\lambda_2)_{\perp}\in   E^{\perp}$.
Therefore, the two conditions under which $\|T_1(\omega_1)-T_1(\omega_2)\|\neq 0$ can be restated as: case 1) $x_1\neq x_2$, and case 2) $(\lambda_1-\lambda_2)_{\perp}\neq 0$.
Next we show that in both cases,   (\ref{eq:co_5}) can be satisfied when $k$ is sufficiently large.

\begin{enumerate}
  \item When $x_1\neq x_2$ is true, the strongly monotone condition in Assumption \ref{ass:cocoersive} implies
  $
  \|T_1(\omega_1)-T_1(\omega_2)\|\geq \mu \|x_1-x_2\|=\mu\|\omega_1-\omega_2\|\frac{\|x_1-x_2\|}{\|\omega_1-\omega_2\|}.
  $
  Because  the primal variables $x_1$ and $x_2$ are always within the compact set $\Omega$,  and the dual variables are bounded (see Proposition \ref{Le:bounded_iterates}), $\|\omega_1-\omega_2\|$ is always bounded, implying  the  existence  of some $k_1$ such that for all $k\geq k_1$  (when $\alpha^k$ is sufficiently small),  $\mu\frac{\|x_1-x_2\|}{\|\omega_1-\omega_2\|}\geq \frac{\alpha^k \|M^k\|}{\frac{\delta}{2}\big(1-2\alpha^k\|M^k\|-(\alpha^k)^2\|M^k\|^2\big)}$ holds (hence (\ref{eq:co_5}) holds) under $x_1\neq x_2$.
  \item When $(\lambda_1-\lambda_2)_{\perp}\neq 0$ is true, using Assumption \ref{ass:cocoersive} and an argument similar to Lemma 3 of \cite{belgioioso2020distributed}, we have\vspace{-0.1cm}
  \[
  \begin{aligned}
  &\|T_1(\omega_1)-T_1(\omega_2)\|\geq \mu \|x_1-x_2\|+\|\Pi_f(\lambda_1-\lambda_2)_{\perp}\|\\
  &\geq\mu \|x_1-x_2\|+{\rm eig}_2(\Pi_f)\|(\lambda_1-\lambda_2)_{\perp}\|\\
  &\geq \textstyle\min\left\{\mu, \frac{{\rm eig}_2(\Pi_f)\|(\lambda_1-\lambda_2)_{\perp}\|}{\|\lambda_1-\lambda_2\|}\right\}(\|x_1-x_2\|+\|\lambda_1-\lambda_2\|)\\
   &\geq \textstyle\min\left\{\mu, \frac{{\rm eig}_2(\Pi_f)\|(\lambda_1-\lambda_2)_{\perp}\|}{\|\lambda_1-\lambda_2\|}\right\} \|\omega_1-\omega_2\|,
  \end{aligned}
  \]
  where
  $\Pi_f\triangleq (I_m-\frac{{\bf 1}{\bf 1}^T}{m})\otimes I_n$ and ${\rm eig}_2(\Pi_f)$ denotes the second smallest eigenvalue of $\Pi_f$ and is equal to 1. Because $(\lambda_1-\lambda_2)_{\perp}\neq 0$,  and $\|\lambda_1-\lambda_2\|$ is always bounded, we can always find some $k_1$ such that for all $k\geq k_1$ (when $\alpha^k$ is sufficiently small), $\min\left\{\mu, \frac{{\rm eig}_2(\Pi_f)\|(\lambda_1-\lambda_2)_{\perp}\|}{\|\lambda_1-\lambda_2\|}\right\}\geq \frac{\alpha^k \|M^k\|}{\frac{\delta}{2}(1-2\alpha^k\|M^k\|-(\alpha^k)^2\|M^k\|^2)}$ holds (hence (\ref{eq:co_5}) holds) under $(\lambda_1-\lambda_2)_{\perp}\neq 0$.
\end{enumerate}

In summary, (\ref{eq:co_5}) always holds for $k\geq k_1$, implying that for all $k\geq k_1$, $(I-\alpha^kM^k)T_1$ is $\frac{\delta}{2}$-cocoercive in the Euclidean norm, and hence,
$(\Phi^k)^{-1}T_1=\alpha^k(I-\alpha^kM^k)T_1$ is  $\frac{\delta}{2\alpha^k}$-cocoercive in the Euclidean norm.
Using Proportion 4.39 of \cite{bauschke2011convex}, we have that ${\rm Id}-(\Phi^k)^{-1}T_1$ is $\frac{\alpha^k}{\delta}$-averaged in the Euclidean norm for all $k\geq k_1$.

 Using a similar argument, we can prove that there exists a $k_2\geq 0$ such that for all $k\geq k_2$, $(\Phi^k)^{-1}T_2$ is maximally monotone in the Euclidean norm since $T_2$ is maximally monotone (see Lemma 3 in \cite{belgioioso2020distributed}). Hence, the  mapping $({\rm Id}-(\Phi^k)^{-1}T_2)^{-1}$ is $\frac{1}{2}$-averaged in the Euclidean norm for all $k\geq k_2$.

  Then, since for  an $a_1$-averaged operator $T_{a}$ and an $a_2$-averaged operator $T_b$, the composition $T_a\circ T_b$ is $\frac{a_1+a_2-2a_1a_2}{1-a_1 a_2}$-averaged (Proposition 4.44 of \cite{bauschke2011convex}),  there exists a $k_0=\max\{k_1,k_2\}$ such that for all $k\geq k_0$, $R_k$  in (\ref{eq:R_k}) is $\frac{1}{2-\frac{2\alpha^k}{\delta}}$-averaged, and  hence, is nonexpansive (Remark 3.43 of \cite{bauschke2011convex}).

            }


 To prove that for every $\{\omega^k\}$, $\lim_{k\rightarrow\infty}\|R_k(\omega^k)-\omega^k\|$ exists almost surely, we use the expression of $(\Phi^k)^{-1}$ in (\ref{eq:Phi_1}).
 Given   $\sum_{k=0}^{\infty}(\alpha^k)^2<\infty$,  it can be easily seen that   $(\Phi^k)^{-1}$ will converge to the zero matrix, which, in turn, implies that $R^k$ will converge to the identity operator. {   Proposition \ref{Le:bounded_iterates} ensures that  $\omega^k$ is always bounded.} Therefore, we always have that $\lim_{k\rightarrow\infty}\|R_k(\omega^k)-\omega^k\|$ exists almost surely for every $\omega^k$. The fact that Condition 4) in Lemma \ref{le:perburbed_fixed_point} can be guaranteed by $\sum_{k=0}^{\infty}\alpha^k=\infty$  follows \cite{zhu2016distributed,yin2011nash}.
\end{proof}

{
\begin{Remark 1}
  Different from  \cite{belgioioso2020distributed}  which proves that  $\Phi^{-1} T_1$ (resp. $\Phi^{-1}T_2$) is cocoercive (resp. maximally monotone) in a $\Phi$-induced norm ($\Phi$ is time-invariant therein), here we prove that $(\Phi^k)^{-1} T_1$ (resp. $(\Phi^k)^{-1} T_2$) is $\frac{\delta}{2\alpha^k}$-cocoercive (resp. maximally  monotone) in the Euclidean norm.  The  results   are reminiscent to  Theorem 5.2 of \cite{mateos2016noise}, which proves that  transformations  that are small perturbations of the identity matrix do not affect cocoercive properties. The difference is that  $\Phi^k$ here is time-varying, and it is close to an $\alpha^k$-scaled  identity matrix rather than  an identity matrix in  \cite{mateos2016noise}.
\end{Remark 1}
}


\subsection{Algorithm 2 is a stochastically-perturbed version of (\ref{eq:iterate_algo3})}\label{se:perturbation_summable}

In this subsection, we prove that Algorithm 2 is a stochastically perturbed version of (\ref{eq:iterate_algo3}) with the perturbation $e(k)$ satisfying the conditions in Lemma \ref{le:perburbed_fixed_point}. To this end, we first prove that the averages of $\sigma_i^k$, $y_i^k$, and $z_i^k$ in Algorithm 2 always converge to $\bar{x}^k$, $\bar{d}^k$, and $\bar\lambda^k$, respectively:
\begin{Lemma 1}\label{Le:average_track}
Under Assumptions \ref{as:L}, \ref{ass:dp-noises-game},  $\bar\sigma^k=\frac{\sum_{i=1}^{m}\sigma_i^k}{m}$, $\bar y^k=\frac{\sum_{i=1}^{m}y_i^k}{m}$, and  $\bar{z}^k=\frac{\sum_{i=1}^{m}z_i^k}{m}$ in Algorithm 2 converge {\it a.s.} to $\bar{x}^k$, $\bar{d}^k=\frac{\sum_{i=1}^{m}d_i^k}{m}$ with $d_i^k$ defined in (\ref{eq:d_i^k}), and  $\bar{\lambda}^k$, respectively, if
$\sum_{k=0}^{\infty}\gamma^k= \infty$ and $\sum_{k=0}^{\infty}(\gamma^k)^2<\infty$ hold.
Moreover,  $\sum_{k=0}^{\infty}\alpha^k\|\bar\sigma^k-\bar{x}^k\|<\infty$, $\sum_{k=0}^{\infty}\alpha^k\|\bar{z}^k-\bar{\lambda}^k\|<\infty$, and $\sum_{k=0}^{\infty}\alpha^k\|\bar{y}^k-\bar{d}^k\|<\infty$ hold {\it a.s.} if $\sum_{k=0}^{\infty}\frac{(\alpha^k)^2}{\gamma^k}<\infty$ holds.
\end{Lemma 1}
\begin{proof}
  The derivation of the first part follows the line of reasoning in Lemma \ref{le:bar_x=bar_r}, and the derivation of the second part follows the last paragraph of the proof of Theorem 1. Hence, we do not include the proof here.
\end{proof}

To show that Algorithm 2 corresponds to a stochastically-perturbed version of (\ref{eq:iterate_algo3}), we rewrite the updates in Algorithm 2 in the following more compact form:
\begin{equation}\label{eq:alg2_compact}
 \begin{aligned}
 \tilde{x}^k&=\Pi_\Omega\left[x^k-\alpha^k\left(F(x^k,\sigma^k)+C_d^Tz^k\right)\right],\\
 \tilde{\lambda}^k&=\Pi_{\mathbb{R}^{mn}_{+}}\left[\lambda^k+{ \alpha}^k\left( y^k-\lambda^k+z^k \right)\right],\\
 x^{k+1}&=x^k+\gamma^k(\tilde{x}^k-x^k),\\
 \lambda^{k+1}&=\lambda^k+\gamma^k(\tilde{\lambda}^k-\lambda^k),
 \end{aligned}
\end{equation}
with
\begin{equation}\label{eq:y}
\begin{aligned}
y^{k+1}=&(1-\gamma^k)y^k+\chi^k(L\otimes I_n)y^k+\chi^k(L^0\otimes I_n)\xi^k\\
&+(2 C_d\tilde{x}^k-C_dx^k-c_d)\\
&-(1-\gamma^k)(2 C_d\tilde{x}^{k-1}-C_dx^{k-1}-c_d),
\end{aligned}
\end{equation}
where $c_d\triangleq {\rm col}(c_1,\cdots,c_m)$.

Recalling that (\ref{eq:alg3_compact}) corresponds to the nonstationary iteration process in (\ref{eq:iterate_algo3}), comparing (\ref{eq:alg2_compact}) and (\ref{eq:alg3_compact}) yields that Algorithm 2 corresponds to the following
stochastically-perturbed nonstationary fixed-point iteration process
\begin{equation}\label{eq:iteration_alg2}
 \omega^{k+1}=\omega^k+\gamma^k\left(R_k(\omega^k)+e^k-\omega^k\right),
\end{equation}
with $e^k\triangleq {\rm col}(e_x^k, e_\lambda^k)$ given by
\begin{equation}\label{eq:e_k}
\begin{aligned}
 e_x^k=& \Pi_\Omega\left[x^k-\alpha^k\left(F(x^k,\sigma^k)+C_d^Tz^k\right)\right]\\
 &\qquad- \Pi_\Omega\left[x^k-\alpha^k\left(F(x^k,\bar{x}_{\bf 1}^k)+C_d^T\bar{\lambda}_{\bf 1}^k\right)\right],\\
 e_{\lambda}^k=&\Pi_{\mathbb{R}^{mn}_{+}}\left[\lambda^k+{ \alpha}^k\left( y^k-\lambda^k+z^k\right)\right]\\
 &\qquad- \Pi_{\mathbb{R}^{mn}_{+}}\left[\lambda^k+{ \alpha}^k\left(\bar{d}^k_{\bf 1}-\lambda^k+\bar{\lambda}^k_{\bf 1}\right)\right].
\end{aligned}
\end{equation}


\subsection{Combine preceding two subsections to prove the final result}
In this subsection, we combine the preceding two subsections to prove that Algorithm 2 can ensure the almost sure convergence of all agents to the GNE.

\begin{Theorem 1}\label{th:game_convergence}
   Under Assumptions \ref{ass:Kset},\ref{ass:cocoersive},\ref{ass:Lipschitz},\ref{as:L},\ref{ass:dp-noises-game}, when the
   nonnegative non-increasing sequence  $\{\alpha^k\}$  satisfies
     \begin{equation}\label{eq:conditions_convergence_Algo2}
   \textstyle\sum_{k=0}^{\infty}\alpha^k=\infty,\,   \sum_{k=0}^{\infty} (\alpha^k)^2  <\infty,
   \end{equation} and
    the nonnegative sequences $\{\gamma^k\}$ and $\{\chi^k\}$ satisfy
   \begin{equation}\label{eq:conditions_convergence_Algo2}
     \sum_{k=0}^{\infty}\chi^k=\infty,   \sum_{k=0}^{\infty}\frac{(\gamma^k)^2}{\chi^k}<\infty,\sum_{k=0}^{\infty}\frac{(\alpha^k)^2}{\gamma^k}<\infty,
   \end{equation}
   then the iterates $x_i^k$ for $i\in[m]$ in Algorithm 2 converge  almost surely to a GNE $x^\ast$ of (\ref{eq:formulation}).
\end{Theorem 1}

\begin{proof}
The basic idea is to leverage Lemma \ref{le:perburbed_fixed_point}. To this end, we characterize $\|e^k\|$, which can be verified to satisfy
\begin{equation}\label{eq:e_decom}
\|e^k\|\leq \|e_x^k\|+\|e_{\lambda}^k\|.
\end{equation}

Using the nonexpansive property of the projection operator, we have the following relationship for $\|e_x^k\|$:
\begin{equation}\label{eq:e_x^k}
\begin{aligned}
 \|e_x^k\|&\leq \left\|x^k-\alpha^k\left(F(x^k,\sigma^k)+C_d^Tz^k\right)\right.\\
 &\left.\qquad -\left(x^k-\alpha^k\left(F(x^k,\bar{x}_{\bf 1}^k)+C_d^T\bar{\lambda}_{\bf 1}^k\right)\right)\right\|\\
 &\leq \alpha^k \| F(x^k,\sigma^k)-F(x^k,\bar{x}_{\bf 1}^k)   \|+ \alpha^k\|C_d^T\|\|z^k-\bar{\lambda}_{\bf 1}^k  \|\\
 &=\alpha^k \| F(x^k,\sigma^k)-F(x^k,\bar{\sigma}_{\bf 1}^k) +F(x^k,\bar{\sigma}_{\bf 1}^k)- F(x^k,\bar{x}_{\bf 1}^k) \|\\
 &\quad + \alpha^k\|C_d^T\|\|z^k-\bar{z}_{\bf 1}^k +\bar{z}_{\bf 1}^k - \bar{\lambda}_{\bf 1}^k\|\\
 &\leq   \alpha^k \tilde{L}\|  \sigma^k-\bar{\sigma}_{\bf 1}^k  \|+ \alpha^k \sqrt{m}\tilde{L}\| \bar{\sigma}^k- \bar{x}^k\|\\
 &\quad + \alpha^k\|C_d^T\|\|z^k-\bar{z}_{\bf 1}^k  \|+\alpha^k\sqrt{m}\|C_d^T\|\|\bar{z}^k - \bar{\lambda}^k\|,
\end{aligned}
\end{equation}
where we have used the Lipschitz condition in Assumption \ref{ass:Lipschitz} in the last inequality.

 Lemma \ref{Le:average_track} implies $\sum_{k=0}^{\infty}\alpha^k\|\bar\sigma^k-\bar{x}^k\|<\infty$ and $\sum_{k=0}^{\infty}\alpha^k\|\bar{z}^k-\bar{\lambda}^k\|<\infty$   {\it a.s.}
Hence, we only need to consider the first and third terms on the right hand side of (\ref{eq:e_x^k}).

According to (\ref{eq:update_algo2}),  the evolutions of $\sigma_i^k$ and $z_i^k$ follow
\[
\begin{aligned}
           \sigma_i^{k+1}                          &=(1-\gamma^k)\sigma_i^k+\chi^k\textstyle\sum_{j\in \mathbb{N}_i}L_{ij}(\sigma_j^k+\zeta_j^k-\sigma_i^k)+ \gamma^k(\tilde{x}_i^k),
\end{aligned}
\]
and
\[
\begin{aligned}
        z_i^{k+1}                  &=(1-\gamma^k)z_i^k+\chi^k\textstyle\sum_{j\in \mathbb{N}_i}L_{ij}(z_j^k+\upsilon_j^k-z_i^k)+ \gamma^k(\tilde{\lambda}_i^k),
\end{aligned}
\]
respectively.
Therefore, given that $x_i^k$, $\tilde{x}_i^k$, $\tilde{\lambda}_i^k$, $\lambda_i^k$ are bounded from Proposition \ref{Le:bounded_iterates},   Theorem \ref{Th:consensus_tracking} implies   $\sigma_i^k\rightarrow \bar{x}^k$ and $z_i^k\rightarrow \bar{\lambda}^k$ almost surely, and the following relations hold {\it a.s.}:
\begin{equation}\label{eq:sigma_summalbe}
\sum_{k=0}^{\infty}\gamma^k\sum_{i=1}^{m}\|\sigma_i^k-\bar{\sigma}^k\|<\infty,\: \sum_{k=0}^{\infty}\gamma^k\sum_{i=1}^{m}\|z_i^k-\bar{z}^k\|<\infty.
\end{equation}
Combining (\ref{eq:e_x^k})   and (\ref{eq:sigma_summalbe}) implies the following relation {\it a.s.}:
\begin{equation}\label{eq:e_x^ksummable}
\sum_{k=0}^{\infty}\gamma^k \|e_x^k\|<\infty.
\end{equation}

Following the same line of derivation, we have
\begin{equation}\label{eq:e_lambda^k}
\begin{aligned}
 \|e_{\lambda}^k\|\leq & \left\|   \lambda^k+{ \alpha}^k\left( y^k-\lambda^k+z^k\right)     \right.\\
 &\left.\qquad-\left( \lambda^k+{ \alpha}^k\left(\bar{d}^k_{\bf 1}-\lambda^k+\bar{\lambda}^k_{\bf 1}\right)  \right) \right\|\\
\leq &{ \alpha}^k \| y^k- \bar{y}_{\bf 1}^k     \|+{ \alpha}^k\|\sqrt{m}\|\bar{y}^k-\bar{d}^k\| + { \alpha}^k  \|z^k-\bar{z}_{\bf 1}^k  \|\\
&\quad +{ \alpha}^k\|\sqrt{m}\|\bar{z}^k-\bar{\lambda}^k\|.
\end{aligned}
\end{equation}

Lemma \ref{Le:average_track} implies  $\sum_{k=0}^{\infty}{ \alpha}^k\|\bar{z}^k-\bar{\lambda}^k\|<\infty$  and $\sum_{k=0}^{\infty}{ \alpha}^k\|\bar{y}^k-\bar{d}^k\|<\infty$  {\it a.s.} under $\sum_{k=0}^{\infty}\frac{({ \alpha}^k)^2}{\gamma^k}<\infty$.
Hence, we only need to consider the first and third terms on the right hand side of (\ref{eq:e_lambda^k}).

According to Theorem \ref{Th:consensus_tracking}, for the dynamics $y_i^k$ in (\ref{eq:y}), if we can prove that $\|(2C_d\tilde{x}^k-C_dx^k-c_d)-(1-\gamma^k)(2C_d\tilde{x}^{k-1}-C_dx^{k-1}-c_d)\|\leq \gamma^k C$ holds for some $C$,  then we have
$\sum_{k=0}^{\infty}\gamma^k\sum_{i=1}^{m}\|y_i^k-\bar{y}^k\|<\infty$ and hence, $\sum_{k=0}^{\infty}{ \alpha}^k\sum_{i=1}^{m}\|y_i^k-\bar{y}^k\|<\infty$.

Noticing $(2 C_d\tilde{x}^k-C_dx^k-c_d)-(1-\gamma^k)(2 C_d\tilde{x}^{k-1}-C_dx^{k-1}-c_d) = 2C_d( \tilde{x}^k- \tilde{x}^{k-1})-C_d(x^k-x^{k-1}) +\gamma^k(2C_d\tilde{x}^{k-1}-C_dx^{k-1}-c_d)$, we have
\begin{equation}\label{eq:y_1}
\begin{aligned}
& \|2 C_d\tilde{x}^k-C_dx^k-c_d -(1-\gamma^k) (2C_d\tilde{x}^{k-1}-C_dx^{k-1}-c_d)\|\\
 &\leq \|2C_d\| \| \tilde{x}^k- \tilde{x}^{k-1}\|+\|C_d\|\|x^k-x^{k-1}\|\\
 &\quad+\gamma^k\|2 C_d\tilde{x}^{k-1}-C_dx^{k-1}-c_d\|\\
 &\leq \|2C_d\| \| \tilde{x}^k- \tilde{x}^{k-1}\|+\gamma^k\|C_d\|\|\tilde{x}^{k-1}-x^{k-1}\|\\
 &\quad+\gamma^k\|2C_d\tilde{x}^{k-1}-C_dx^{k-1}-c_d\|,
 \end{aligned}
\end{equation}
where we have used the update rule of $x_i^k$ in (\ref{eq:update_algo2}) in the last inequality.

To characterize $\| \tilde{x}^k- \tilde{x}^{k-1}\|$, we employ the update rule of $\tilde{x}_i^k$ in (\ref{eq:update_algo2}). Using the nonexpansive property of projection operators, we have
\begin{equation}\label{eq:y2}
 \begin{aligned}
 &\| \tilde{x}^k- \tilde{x}^{k-1}\|= \left\| \Pi_\Omega\left[x^k-\alpha^k\left(F(x^k,\sigma^k)+C_d^Tz^k\right)\right]\right.\\
 &\left.\qquad -\Pi_\Omega\left[x^{k-1}-\alpha^{k-1}\left(F(x^{k-1},\sigma^{k-1})+C_d^Tz^{k-1}\right)\right] \right\|\\
 &\leq \left\|  x^k-\alpha^k\left(F(x^k,\sigma^k)+C_d^Tz^k\right) \right.\\
 &\left.\qquad - \left(x^{k-1}-\alpha^{k-1}\left(F(x^{k-1},\sigma^{k-1})+C_d^Tz^{k-1}\right)\right) \right\|\\
 &  \leq \|  x^k-x^{k-1}\|+ \alpha^k\left\|F(x^k,\sigma^k)+C_d^Tz^k\right\| \\
 & \qquad +\alpha^{k-1}\left\|F(x^{k-1},\sigma^{k-1})+C_d^Tz^{k-1}\right\|.
 \end{aligned}
\end{equation}

Using the proven results that $\sigma_i^k\rightarrow \bar{x}^k$ and $z_i^k\rightarrow \bar{\lambda}^k$ hold almost surely, and $\bar{x}^k$ and $\bar{\lambda}^k$ are bounded  (see Proposition \ref{Le:bounded_iterates}), we have that
  $z_i^k$ and $\sigma_i^k$ are bounded almost surely. Therefore, the term  $  F(x^k,\sigma^k)+C_d^Tz^k   $ is bounded almost surely. Without loss of generality, we assume
  $
  \|F(x^k,\sigma^k)+C_d^Tz^k \|\leq\bar{C}
  $
   for all $k$ almost surely.

   Plugging the preceding relationship into (\ref{eq:y2}) leads to
\begin{equation}\label{eq:y3}
 \begin{aligned}
 &\| \tilde{x}^k- \tilde{x}^{k-1}\|  \leq \|  x^k-x^{k-1}\|+ (\alpha^k +\alpha^{k-1})\bar{C} \\
 &\leq \gamma^k\|\tilde{x}^k-x^k\|+(\alpha^k +\alpha^{k-1})\bar{C},
 \end{aligned}
\end{equation}
where we have used the update rule of $x_i^k$ in (\ref{eq:update_algo2}) in the last inequality.
Given $\sum_{k=0}^{\infty}\frac{(\alpha^k)^2}{\gamma^k}<\infty$ in the theory statement, we have  $(\alpha^k +\alpha^{k-1})\leq \tilde{C}\gamma^k$ for some constant $\tilde{C}$, and hence
$\| \tilde{x}^k- \tilde{x}^{k-1}\|
 \leq \gamma^k (\|\tilde{x}^k-x^k\|+ \tilde{C})
$ and further
\begin{equation}\label{eq:y5}
 \begin{aligned}
&\|C_d(2\tilde{x}^k-x^k-c_d)-(1-\gamma^k)C_d(2\tilde{x}^{k-1}-x^{k-1}-c_d)\|\\
&\leq  \gamma^k\|C_d\|\left(\|\tilde{x}^{k-1}-x^{k-1}\|+2\|\tilde{x}^k-x^k\|+ 2\tilde{C}\|\right)\\
&\quad+\gamma^k\| 2 C_d\tilde{x}^{k-1}-C_dx^{k-1}-c_d\|
 \end{aligned}
\end{equation}
by using (\ref{eq:y_1}).

Combining (\ref{eq:y}) and (\ref{eq:y5}), we have that the evolution of $y_i^k$ satisfies the conditions of Theorem \ref{Th:consensus_tracking}, and hence, $\sum_{k=0}^{\infty}\gamma^k\sum_{i=1}^{m}\|y_i^k-\bar{y}^k\|<\infty$ holds almost surely. Further invoking (\ref{eq:sigma_summalbe}) and (\ref{eq:e_lambda^k}) yields the following result {\it a.s.}:
\begin{equation}\label{eq:e_lambda^ksummable}
\sum_{k=0}^{\infty}\gamma^k \|e_{\lambda}^k\|<\infty.
\end{equation}

 In summary, (\ref{eq:e_decom}), (\ref{eq:e_x^ksummable}), and (\ref{eq:e_lambda^ksummable}) mean that $\sum_{k=0}^{\infty}\gamma^k \|e^k\|<\infty$ holds   under the theorem statement almost surely. {  Following the arguments in \cite{zhu2016distributed,yin2011nash}, under $\sum_{k=0}^{\infty}\alpha^k=\infty$, this implies that the Condition 4) in Lemma \ref{le:perburbed_fixed_point} will also be guaranteed almost surely for the perturbed algorithm 2.} Moreover, the condition $\sum_{k=0}^{\infty}\gamma^k \|e^k\|<\infty$ means $\sum_{k=0}^{\infty}\gamma^k \mathbb{E}[\|e^k\|]<\infty$   almost surely. Given $\sqrt{\mathbb{E}[\|e^k\|^2 ]}\leq \mathbb{E}[\|e^k\|]$ due to  Jensen's inequality,  invoking   Lemma \ref{le:perburbed_fixed_point} yields that Algorithm 2 guarantees the convergence of all players to the GNE almost surely.
\end{proof}

\section{Privacy analysis of Algorithm 2}\label{se:DP_algo2}

Similar to Definition \ref{de:sensitivity}, we define the sensitivity of a distributed GNE seeking algorithm to problem (\ref{eq:formulation}) as follows:
\begin{Definition 1}\label{de:sensitivity_game}
  At each iteration $k$, for any initial state $\vartheta^0$ and any adjacent distributed GNE problems  $\mathcal{P}$ and $\mathcal{P'}$,  the sensitivity of a GNE seeking algorithm is
    \begin{equation}\label{eq:sensitivity_algo2}
  \Delta^k\triangleq \sup\limits_{\mathcal{O}\in\mathbb{O}}\left\{\sup\limits_{\Theta\in\mathcal{R}_{\mathcal{P},\vartheta^0}^{-1}(\mathcal{O}),\:\Theta'\in\mathcal{R}_{\mathcal{P'},\vartheta^0}^{-1}(\mathcal{O})}\hspace{-0.3cm}\|\Theta^{k}-\Theta'^{k}\|_1\right\},
  \end{equation}
  where $\Theta^k={\rm col}(\sigma^k,y^k,z^k)$.
\end{Definition 1}

Then, similar to Lemma \ref{Le:Laplacian}, we have the following lemma:
\begin{Lemma 1}\label{Le:Laplacian_game}
In Algorithm 2, at each iteration $k$, if each player adds noise vectors $\zeta_i^k$,   $\upsilon_i^k$, and $\xi_i^k$, to its shared messages $\sigma_i^k$, $z_i^k$, and $y_i^k$, respectively, with every noise vector consisting of  independent Laplace scalar noises with  parameter $\nu^k$, such that $\sum_{k=1}^{ \bar{k}}\frac{\Delta^k}{\nu^k}\leq \bar\epsilon$, then  Algorithm 2 is ${ \bar\epsilon}$-differentially private with the cumulative privacy budget for iterations from $k=1$ to $k= \bar{k}$ less than $\bar\epsilon$.
\end{Lemma 1}
\begin{proof}
The lemma can be obtained following the same line of reasoning of Lemma 2 in  \cite{Huang15} (also see Theorem 3 in \cite{ye2021differentially}). Please also see the explanations  in the proof of Lemma \ref{Le:Laplacian}.
\end{proof}

\begin{Theorem 1}\label{th:DP_Algorithm2}
{ Under the conditions of Theorem \ref{th:game_convergence}},  if all elements of $\zeta_i^k$, $\xi_i^k$, and $\upsilon_i^k$ are drawn independently from  Laplace distribution ${\rm Lap}(\nu^k)$ with $(\sigma_i^k)^2=2(\nu^k)^2$ satisfying Assumption \ref{ass:dp-noises-game}, then all players in Algorithm 2
will converge almost surely to the GNE. Moreover,  
\begin{enumerate}
\item For any finite number of iterations $ \bar{k}$, Algorithm 1 is  $\epsilon$-differentially private with the cumulative privacy budget bounded by $\epsilon\leq \sum_{k=1}^{ \bar{k}}\frac{ (\varsigma_y^k+\varsigma_\sigma^k+\varsigma_z^k)}{\nu^k}$   where  $\varsigma_y^k\triangleq C_y(\sum_{p=1}^{k-1}(\Pi_{q=p}^{k-1}(1-\gamma^{q}-\bar{L}\chi^{q})(2-\gamma^{p-1}))+2-\gamma^{k-1})$, $\varsigma_\sigma^k\triangleq C_\sigma(\sum_{p=1}^{k-1}(\Pi_{q=p}^{k-1}(1-\gamma^q-\bar{L}\chi^{q})) \gamma^{p-1}+ \gamma^{k-1})$, $\varsigma_z^k\triangleq C_z(\sum_{p=1}^{k-1}(\Pi_{q=p}^{k-1}(1-\gamma^q-\bar{L}\chi^{q})) \gamma^{p-1}+ \gamma^{k-1})$,  $\bar{L}\triangleq\min_i\{|L_{ii}|\}$, $C_y \triangleq \max_{i\in[m],0\leq k\leq  \bar{k}}\|  {d}^k_i - {d_i'}^k  \|_1$, $C_\sigma \triangleq \max_{i\in[m],0\leq k\leq  \bar{k}}\|\tilde{x}^k_i -(\tilde{x}_i^k)'\|_1$, $C_z \triangleq \max_{i\in[m],0\leq k\leq \bar{k}}\|\tilde{\lambda}^k_i -(\tilde{\lambda}_i^k)'\|_1$;
\item  The cumulative privacy budget is  finite for $ \bar{k}\rightarrow\infty$  when the sequence  $\{\frac{ \gamma^k }{\nu^k}\}$ is summable.
\end{enumerate}
\end{Theorem 1}
\begin{proof}
Since the Laplace noises satisfy Assumption \ref{ass:dp-noises-game}, the convergence follows directly from Theorem 3.

According to the definition of sensitivity in Definition \ref{de:sensitivity_game}, we can obtain $\Delta^k=\Delta^k_\sigma+\Delta^k_y+\Delta^k_z$, where $\Delta^k_\sigma$, $\Delta^k_y$, and $\Delta^k_z$ are obtained by replacing $\Theta^k$ in (\ref{eq:sensitivity_algo2}) with $\sigma^k$, $y^k$, and $z^k$, respectively (note that the norm is $L_1$ norm).

Given two adjacent networked games  $\mathcal{P}$ and $\mathcal{P'}$,
 we  represent the different cost functions as $J_i$ in $\mathcal{P}$ and $J'_i$ in $\mathcal{P}'$  without loss of generality.

 Here, we only derive the result for $\Delta^k_\sigma$, but $\Delta^k_y$ and $\Delta^k_z$ can be obtained using the same argument.

 Because the initial conditions, cost functions, and observations of $\mathcal{P}$ and $\mathcal{P'}$  are identical for $j\neq i$, we have $\sigma_j^k={\sigma'_j}^k$ for all $j\neq i$ and $k$. Therefore, $\|\sigma^{k}-\sigma'^{k}\|_1$ is always equal to $\|\sigma_i^{k}-{\sigma'_i}^{k}\|_1$.

According to   Algorithm  2, we can arrive at
 \[
 \begin{aligned}
 \|\sigma_i^{k+1}-{\sigma'_i}^{k+1}\|_1\leq& (1-\gamma^k-|L_{ii}|\chi^k)\|\sigma_i^k-{\sigma'_i}^k\|\\
 &+\gamma^k\|\tilde{x}_i^k- (\tilde{x}_i^k)'\|_1,
 \end{aligned}
 \]
 where we have   used the  relationship $x_i^{k+1}-(1-\gamma^k )x_i^k=\gamma^k\tilde{x}_i^k$ and the fact that the observations $\sigma_j^k+\sigma_j^k$ and ${\sigma'_j}^k+{\sigma'_j}^k$ are the same.

 Hence,  $\Delta_\sigma^k$ satisfies
 $
 \Delta_\sigma^{k+1}\leq (1-\gamma^k-|L_{ii}|\chi^k)\Delta_\sigma^{k}+\gamma^k\|\tilde{x}_i^k- (\tilde{x}_i^k)'\|_1$.

Using a similar line of argument, we can obtain the iteration relations for $\Delta^k_y$ and $\Delta^k_z$, and hence, arrive at the first privacy statement by iteration. 

 For the infinite-horizon result in the second privacy statement, we exploit the fact that   our algorithm ensures convergence of both $\mathcal{P}$ and $\mathcal{P'}$ to their respective GNE points, which are the same under the third requirement in Definition \ref{de:adjacency}. This means that  $\|\tilde{x}_i^k- (\tilde{x}_i^k)'\|_1=0$, $\|\tilde{\lambda}_i^k- (\tilde{\lambda}_i^k)'\|_1=0$, and $\|d_i^k-  {{d}_i'}^k\|_1=0$  will hold when $k$ is large enough. Furthermore, the ensured convergence also means that $\|\tilde{x}_i^k- (\tilde{x}_i^k)'\|_1$, $\|\tilde{\lambda}_i^k- (\tilde{\lambda}_i^k)'\|_1$, and $\|d_i^k-  {{d}_i'}^k\|_1$ are always  bounded. Hence, there always exists some constant ${C}$ such that the sequences $\{\|\tilde{x}_i^k- (\tilde{x}_i^k)'\|_1\}$, $\{\|\tilde{\lambda}_i^k- (\tilde{\lambda}_i^k)'\|_1\}$, and $\{\|d_i^k-  {{d}_i'}^k\|_1\}$   are upper bounded by the sequence $\{ {C}\chi^k \gamma^k \}$.

 Therefore, according to Lemma \ref{le:chung},   there always exists a constant $\bar{C}$ such that  $\Delta^k\leq \bar{C} \gamma^k  $ holds (note that $\{\gamma^k\}$ decays faster than $\{\chi^k\}$).
Using Lemma \ref{Le:Laplacian}, we can easily obtain $\epsilon\leq \sum_{k=1}^{ \bar{k}}\frac{\bar{C} \gamma^k }{\nu^k}$. Hence, $\epsilon$ will  be finite even when $ \bar{k}$ tends to infinity if  the sequence  $\{\frac{\gamma^k}{\nu^k}\}$ is summable, i.e.,  $\sum_{k=0}^{\infty}\frac{ \gamma^k }{\nu^k}<\infty$.
\end{proof}


\begin{Remark 1}
  {  Similar to the consensus-tracking case,} to ensure that the  cumulative privacy budget $\epsilon=\sum_{k=1}^{\infty}\frac{\bar{C}\gamma^k}{\nu^k}$ is bounded when  $k\rightarrow \infty$, our algorithm uses   Laplace noise with parameter $\nu^k$ that increases with time (since we require  $\{\frac{\gamma^k}{\nu^k}\}$ to be summable while  $\{\gamma^k\}$ is non-summable).  In addition, according to Theorem \ref{th:game_convergence}, the convergence is not affected by scaling  $\nu^k$ by any constant coefficient $\frac{1}{\epsilon}>0$  to achieve any desired level of $\epsilon$-DP, as long as the DP-noise parameter $\nu^k$  satisfies Assumption \ref{ass:dp-noises-game}. 
\end{Remark 1}

\section{Numerical Simulations}\label{se:simulation}
\begin{figure}
\center
\includegraphics[width=0.45\textwidth]{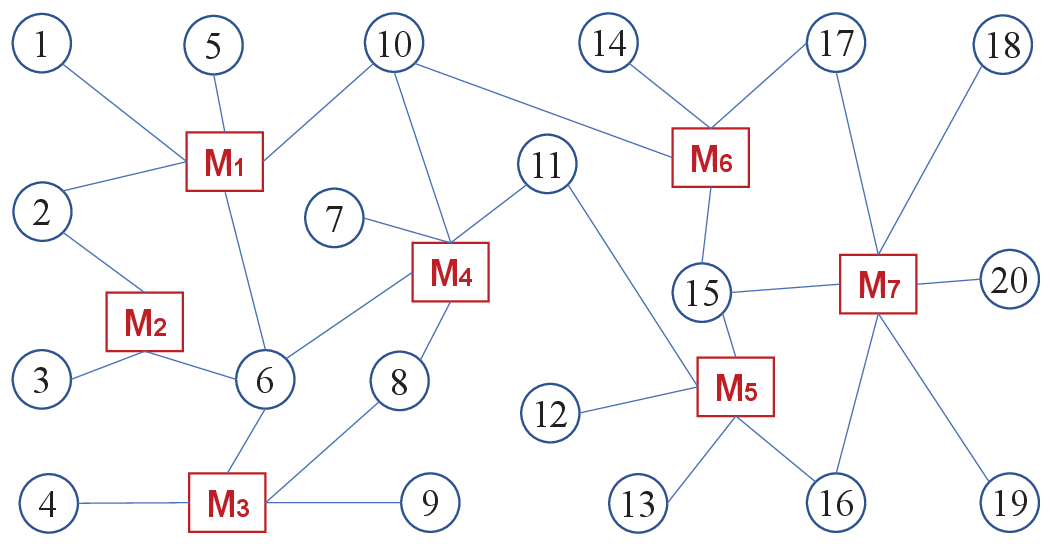}
    \caption{Nash-Cournot game of 20 players (firms) competing over 7 locations (markets). Each firm is represented by a circular and each market is represented by a square. An edge between firm $i$ $(1\leq i\leq 20)$ and market $j$ ($1\leq j\leq 7$) means that firm $i$ participates in market $j$. }
    \label{fig:network}
\end{figure}

 We evaluate the performance of the  proposed GNE seeking algorithm   using a Nash-Cournot game  recently considered in    \cite{pavel2019distributed,koshal2016distributed,nguyen2022distributed}. In the game, we consider   $m$ firms  producing a homogeneous commodity competing over $N$ markets, with a schematic presented in  Fig. \ref{fig:network}. In the schematic,  we plot $N=7$ markets (represented by $M_1,\,\cdots,M_7$) and $m=20$ firms (represented by circles). We use an edge from circle $i$ to $M_j$ to denote  that firm $i$ participates in market $M_j$.

 We  use  $x_i\in\mathbb{R}^N$ to represent the amount of firm $i$'  products. If firm $i$ does not participate in   market $j$, then the $j$th entry of $x_i$ will be forced to be $0$ all the time. Hence, if  firm $i$ participates in  $1\leq n_i\leq N$ markets, then its production vector $x_i$ will have $n_i$ non-zero entries. For notational simplicity, we use an adjacency matrix $B_i\in\mathbb{R}^{N\times N}$  to describe the association relationship between firm $i$ and all  markets. More specifically, $B_i$'s  off-diagonal elements are all zero, and its  $j$th diagonal entry is one if firm $i$ participates in market $j$, otherwise, its $j$th diagonal entry is zero. Every firm $i$ has a maximal capacity for each market $j$ it participates in, which is represented by $C_{ij}$. Defining $\bar{C}_i\triangleq [C_{i1},\,\cdots,\,C_{iN}]^T$, the capacity constraint can be formulated as $x_i\leq \bar{C}_i$.
 Defining $B\triangleq [B_1,\,\cdots,\,B_N]$, $Bx\in\mathbb{R}^N=\sum_{i=1}^{N}B_ix_i$ represents the total product supply to all markets, given firm $i$'s production amount $x_i$. We assume that every market has a maximum capacity $\bar{c}_i$. Defining $\bar{c}\triangleq [\bar{c}_1,\,\cdots,\,\bar{c}_N]^T\in \mathbb{R}^N$, the shared coupling constraints can be formulated as $Bx\leq \bar{c}$.

 As in \cite{pavel2019distributed}, we let the commodity's price in every  market $M_i$  follow  a linear inverse demand function:
$
 p_i(x)=\bar{P}_i-\varsigma_i[Bx]_i,$
 where $\bar{P}_i$ and $\varsigma_i>0$ are  constants and $[Bx]_i$ denotes the $i$th element of the vector $Bx$. It can be verified that the price value decreases when the amount of supplied commodity  increases.

Representing the price vector of all markets as $p \triangleq [p_1,\,\cdots,\,p_N]^T$, we have
 $
 p=\bar{P}-\Xi Bx,$
 where $\bar{P}\triangleq[\bar{P}_1 ,\cdots,\bar{P}_N]^T$ and $\Xi\triangleq{\rm diag}(\varsigma_1,\,\cdots,\,\varsigma_N)$. The total payoff of firm $i$ can then be expressed as $p^TB_ix_i$. Firm $i$'s production cost is assumed to be a quadratic function
$
 \phi_i(x_i)=x_i^TQ_ix_i+q_i^Tx_i,$
 where $Q_i\in\mathbb{R}^{N\times N}$ is  positive definite  and $q_i\in\mathbb{R}^{N}$.

Firm $i$'s local cost function, which is determined by its production cost $\phi_i$ and payoff, is given by
$
J_i(x_i,x)=\phi_i(x_i)-(\bar{P}-\Xi Bx)^T B_i^T x_i$.
 One can verify that the gradient of the cost function is
 $
 F_i(x_i,x)=2Q_ix_i+q_i+B_i^T\Xi B_ix_i-B_i(\bar{P}-\Xi B x).$
 It is clear that both firm $i$'s local cost function and gradient are dependent on other firms' actions.

\begin{figure}
\center
\includegraphics[width=0.45\textwidth]{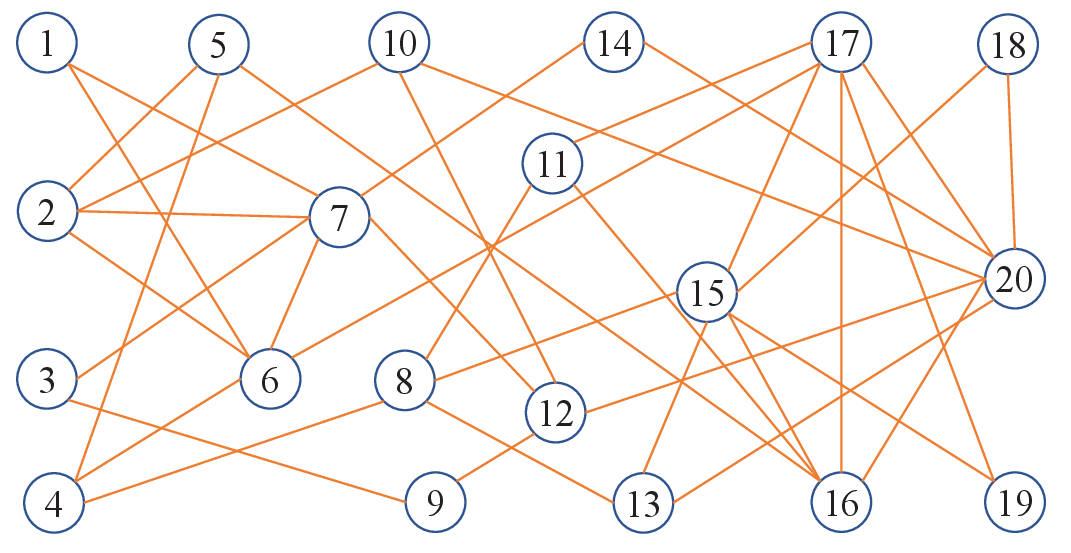}
    \caption{The randomly generated interaction patten of the 20 firms.}
    \label{fig:topology}
\end{figure}

 In our simulation, we consider $m=20$ firms competing over $N=7$ markets. Since in the partial-decision information setting, each firm can only communicate with its immediate neighboring firms, we use a randomly generated  local communication   pattern  given in Fig. \ref{fig:topology}. The   maximal capacities for firm $i$ (elements in $\bar{C}_i$) are randomly selected from the interval $[8,\,10]$. The maximal capacities for all markets are set as $\kappa\sum_{i=1}^{20}\bar{C}_i$, with $\kappa$ randomly selected from a uniform distribution on $(0,\,1)$. $Q_i$ in the production cost function is set as $\nu I$ with $\nu$ randomly selected from $[1,10]$. $q_i$ in $\phi_i(x_i)$ is randomly selected from a uniform distribution in $[1,\,2]$. In the price function, $\bar{P}_i$ and $\varsigma_i$ are randomly chosen from uniform distributions in $[10,\,20]$ and $[1,\,3]$, respectively.

\begin{figure}
\center
\includegraphics[width=0.45\textwidth]{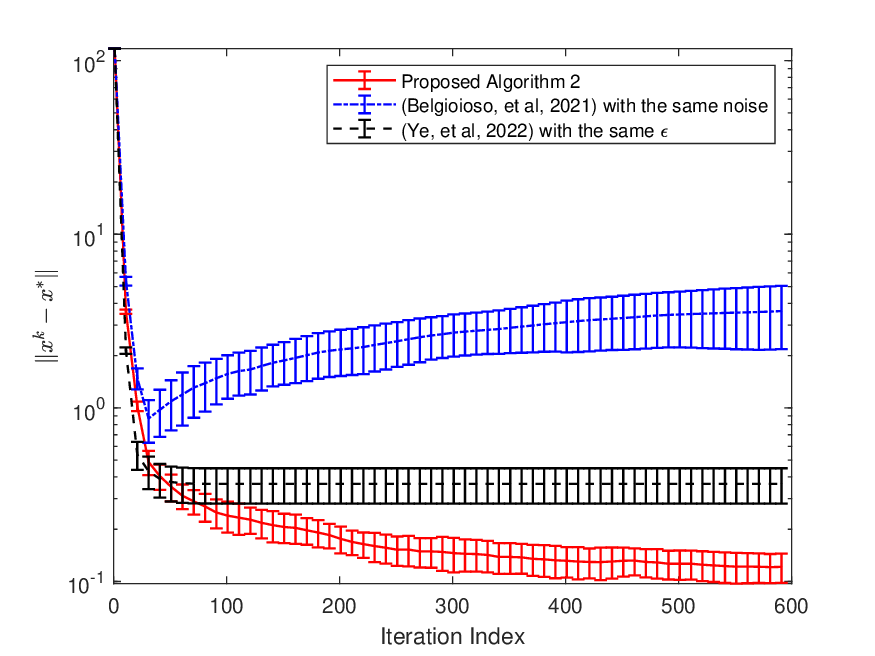}
    \caption{Comparison of Algorithm 2 with the existing GNE seeking algorithm by Belgioioso et al. in \cite{belgioioso2020distributed} (under the same noise) and the  differential-privacy approach  from Ye et al. in \cite{ye2021differentially} after adaptation to GNE seeking (under the same privacy budget $\epsilon$).}
    \label{fig:comparison_algo1}
\end{figure}

  To evaluate the proposed Algorithm 2, for every firm $i$, we  inject  DP-noises $\zeta_i^k$, $\xi_i^k$, and $\upsilon_i^k$ in every shared $\sigma_i^k$, $y_i^k$, and $z_i^k$   in all iterations. Each element of the noise vectors follows Laplace  distribution with parameter $\nu^k=1+0.1k^{0.2}$.  We set the stepsizes and diminishing sequences as   $\alpha^k=\frac{0.1}{1+0.1k}$, $\gamma^k =\frac{0.01}{1+0.1k^{0.98}}$ and $\chi^k =\frac{1}{1+0.1k^{0.9}}$, respectively, which satisfy the conditions in Theorem 3. In the evaluation, we run our algorithm for 100 times, and calculate the average and the variance of  $\|x^k-x^{\ast}\|$ against the iteration index $k$. The result is given by the red curve and error bars in Fig. \ref{fig:comparison_algo1}. For comparison, we also run the existing GNE seeking algorithm  proposed by Belgioioso et al. in \cite{belgioioso2020distributed} under the same noise, and the existing DP approach for Nash equilibrium seeking proposed by Ye et al. in \cite{ye2021differentially} under the same cumulative privacy budget $\epsilon$. Note that the DP approach in \cite{ye2021differentially} addresses  Nash equilibrium seeking problems without shared coupling constraints. We adapt  its DP mechanism (geometrically decreasing stepsizes and DP-noises) to the GNE seeking problem.   The evolution of the average error/variance of the  approaches in \cite{belgioioso2020distributed}  and \cite{ye2021differentially} are given by  the blue and black curves/error bars in Fig. \ref{fig:comparison_algo1}, respectively. It can be seen that the proposed algorithm has a   much better  accuracy.

\section{Conclusions}\label{se:conclusions}

This paper proposes a differentially-private fully distributed algorithm for generalized Nash equilibrium seeking. Different from  existing privacy solutions for  coupling-constraint free aggregative games, the proposed approach allows the existence of shared coupling constraints, which  increase  attack surfaces, and hence, pose additional challenges to privacy protection.  More interestingly, the proposed approach can    ensure both provable convergence to the generalized Nash equilibrium point and rigorous $\epsilon$-differential privacy.  As a basis of the differentially private generalized Nash equilibrium seeking algorithm, we also propose a new consensus-tracking algorithm that can ensure both provable convergence accuracy and rigorous $\epsilon$-differential privacy. The convergence analysis generalizes existing results of stochastically-perturbed nonstationary fixed-point iteration processes to the diminishing-stepsize case, which is crucial to ensure a  finite privacy budget, and hence, rigorous differential privacy.     Numerical simulation   results   confirm the effectiveness of the proposed algorithm.


\bibliographystyle{IEEEtran}

\bibliography{reference1}

\end{document}